\documentclass[12pt,a4paper,twoside,openright]{book}
\usepackage{a4wide}
\usepackage{amsmath,amsfonts,amssymb,latexsym}
\usepackage{epic,eepic}
\usepackage[cp850]{inputenc}
\usepackage{setspace}
\usepackage{graphicx}
\usepackage{epsfig}
\usepackage[dvips]{lscape}
\usepackage{fancyhdr}
\renewcommand{\headrulewidth}{0.4pt}
\newcommand{\chshort}{} 
\newcommand{\Date}{}
\newcommand{\Revision}{}
\newcommand{\chset}[1]{\renewcommand{\chshort}{\thechapter. #1}}

\makeatletter
\def\cleardoublepage{\clearpage\if@twoside \ifodd\c@page\else
 \hbox{}
 \vspace*{\fill}
 \thispagestyle{empty}
 \newpage\fi\fi}
\makeatother
\usepackage{cite}
\usepackage[small]{caption}
\def \be{\begin{equation}}
\def \ber{\begin{eqnarray}}
\def \ee{\end{equation}}
\def \eer{\end{eqnarray}}

\begin{document}
\pagenumbering{roman}
\fancyhf{} 
\fancyhead[CE]{M. H Szymanska --- Bose condensation and lasing in optical microstructures}
\fancyhead[LO]{\sl \leftmark}
\fancyfoot[CO,CE]{\thepage}
\fancypagestyle{plain}{
\renewcommand{\headrulewidth}{0pt}
\fancyhf{}
\fancyfoot[C]{\thepage}
\fancyfoot[L]{\Date}
\fancyfoot[R]{\Revision}}
\thispagestyle{plain}
\frontmatter
\begin{center}
\Large
\vspace*{\stretch{1}}
\Huge
\textbf{Bose Condensation and Lasing\\
        in Optical Microstructures}

\vspace*{\stretch{1}}

\textbf{ Part 2}  \\[.5cm]

\vspace*{\stretch{1}}
\normalsize

\large

\textbf{ Marzena Hanna Szymanska}  \\[.5cm]

\normalsize
 Trinity College \\
University of Cambridge

\vspace*{\stretch{1}}
\normalsize

\begin{center}
\includegraphics[width=0.16\linewidth,angle=0]{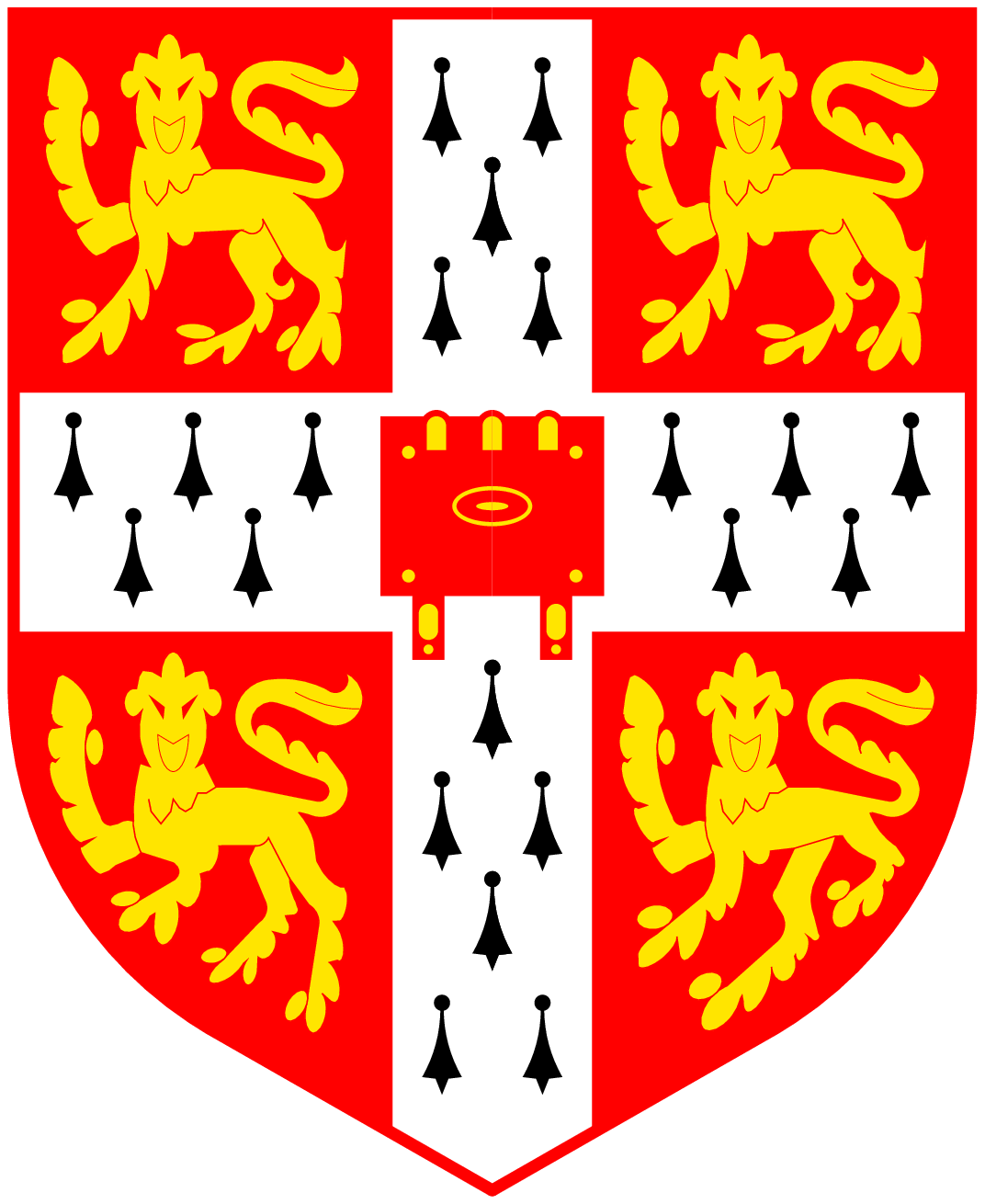}
\end{center}

\vspace*{\stretch{1}}
\normalsize

Dissertation submitted for the degree \\
of Doctor of Philosophy at the \\
University of Cambridge
\\[.2cm] 
{\em October 2001}

\end{center}

\tableofcontents
\cleardoublepage
\mainmatter
\pagenumbering{arabic}
\fancyhf{} 
\fancyhead[CE]{M. H Szymanska --- Bose Condensation and Lasing in Optical Microstructures }
\fancyhead[LO]{\chshort}
\fancyhead[RO,LE]{\thepage}
\fancyfoot[L]{\Date}
\fancyfoot[R]{\Revision}
\fancypagestyle{plain}{
\renewcommand{\headrulewidth}{0pt}
\fancyhf{}
\fancyfoot[C]{\thepage}
\fancyfoot[L]{\Date}
\fancyfoot[R]{\Revision}}
\part{Excitons in T-shaped quantum wires}

\chapter{Introduction}
\chset{Introduction}
\label{Intr}

Optical properties of electrons and holes confined to few dimensions
are of interest for optical and electronic devices.  As the
dimensionality of the structure is reduced, the density of states
tends to bunch together leading to a singularity in the 1D case. This
effect can be very useful for low-threshold laser applications.  At
the same time the excitonic interaction in 1D is enhanced with respect
to that in 3D and 2D structures. Quantum confinement leads to an
increase in the exciton binding energy, $E_b$, and the oscillator
strength for radiative recombination.  Both effects provide
possibilities for much better performance of optical devices such as
semiconductor lasers.
                       
The binding energy of a ground-state exciton in an ideal 2D quantum
well is four times that in the 3D bulk semiconductor.  For the ideal
1D quantum wire $E_b$ diverges. This suggests that $E_b$ for quasi-1D
wires can be greatly increased with respect to the 2D limit for very
thin wires with high potential barriers. 3D and 2D excitons dissociate
at room temperature in GaAs to form an electron-hole plasma. To make
them useful for real device applications, their binding energy needs
to be increased and this might be achieved by using 1D quantum
confinement.

Technologically it is very difficult to manufacture good quality 1D
quantum wires with confinement in both spatial directions. They can be
obtained from a 2D quantum well, fabricated by thin-film growth, by
lateral structuring using lithographic methods. The accuracy of this
method is, however, limited to some ten nanometers and thus the
electronic properties of samples constructed in this way typically
have a strong inhomogeneous broadening. Fortunately it appears
possible to achieve quasi-1D particles even without a rigorous
confinement in any of the spatial directions. This has been realised
in so called V and T-shaped quantum wires. V-shaped quantum wires are
obtained by self-organised growth in pre-patterned materials such
as chemically etched V-shaped grooves in GaAs substrates. The T-shaped
quantum wire, first proposed by Chang et al.\ \cite{Chang}, forms at
the intersection of two quantum wells and is obtained by the cleaved
edge over-growth (CEO) method, a molecular-beam epitaxy (MBE)
technique. The accuracy of this method is extremely high and allows
fabrication of very thin (less than the Bohr radius of an exciton)
wires with small thickness fluctuations. These structures are
currently the subject of intensive research and have been realised by
several groups \cite{N2N4}-\cite{R}.

Experimentalists try to optimise the geometry and the materials in
order to increase the binding energy of the excitons, $E_b$, and the
confinement energy, $E_{con}$ for possible room temperature
applications. Up until now, the most popular material studied
experimentally has been GaAs/Al$_{x}$Ga$_{1-x}$As. Increasing the Al
molar fraction, $x$, should lead to bigger $E_b$ and $E_{con}$ but,
unfortunately, for larger $x$ the interfaces get rougher which
degrades the transport properties. Thus optimised geometries for lower
values of $x$ become more relevant.

The confinement energy, $E_{con}$, is the energy difference between
the lowest excitonic state in the wire and the lowest excitonic state
in the 2D quantum well. It can be directly measured as the difference
between the photoluminescence peaks obtained in a quantum wire (QWR)
and a quantum well (QW). It is, however, not possible to measure the
exciton binding energy directly. Its value has to be obtained from a
combination of experimental data and one-particle calculations of
electron and hole energies in a wire. There has been a disagreement
between the purely theoretical values \cite{Nonvar1}--\cite{Var3} and
those obtained from a combination of experimental data and theoretical
calculations.  The confinement energies, however, tend to agree
between experiment and purely theoretical calculations, suggesting
that experiment, using combined methods where errors tend to
accumulate, usually overestimates the binding energy.

\begin{figure}[htbp]
      \begin{center}
        \leavevmode
      \epsfxsize=8.0cm
      \epsfbox{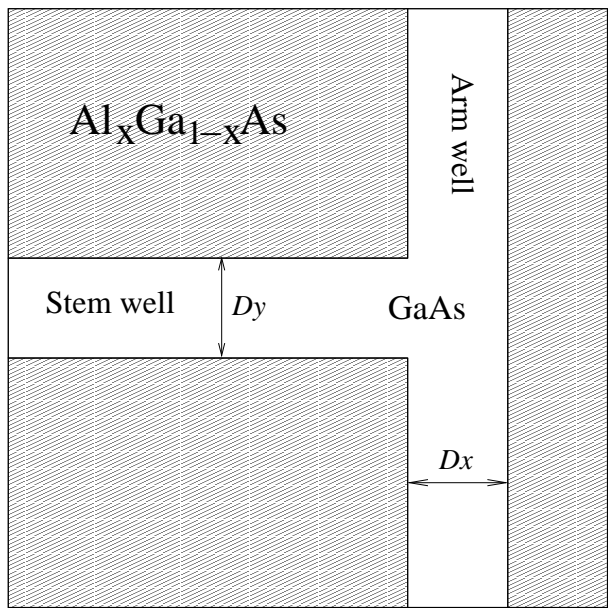}
        \end{center}
  \caption{Shape of the symmetric T-shaped wire with notations. }
\label{sympot}
\end{figure}

\begin{figure}[htbp]
      \begin{center}
        \leavevmode
      \epsfxsize=8.0cm
      \epsfbox{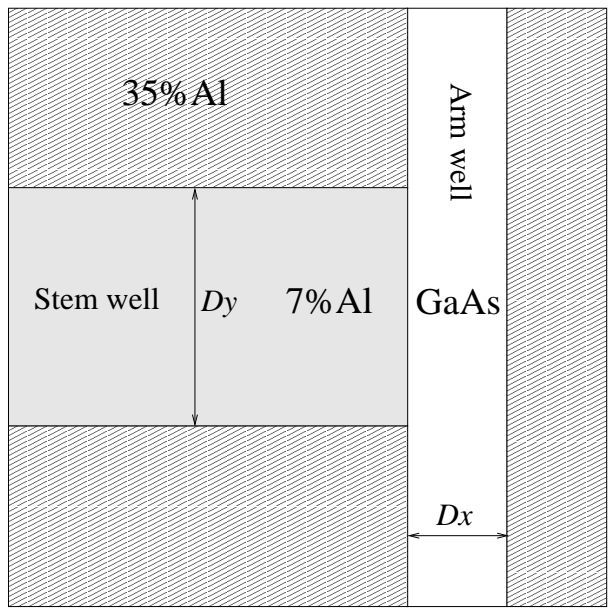}
        \end{center}
  \caption{Shape of the asymmetric T-shaped wire with notations. }
\label{asympot}
\end{figure}

For the 5-nm scale symmetric GaAs/AlAs, Someya et al. \cite{S1S2}
reported the largest confinement energy for excitons in symmetric
wires (Figure \ref{sympot}), $E_{con}$=38 meV and $E_b$=27$\pm$3
meV. 
The largest confinement energy of any structure was reported by
Gislason et al. \cite{G1,G2} for their optimised wires. Using
asymmetric wells with different widths and Al content as in
Figure \ref{asympot}, they obtained an exciton confinement energy of 54
meV. Recently there has also been the first experimental realisation
of T-shaped wires using In$_{y}$Ga$_{1-y}$As/Al$_{0.3}$Ga$_{0.7}$As
\cite{N2N4}. The highest confinement energy reported for this
structure is 34 meV, which is very close to the GaAs/AlAs result, and
the quality of the structure can be much higher than for the GaAs/AlAs
case.

Laser emission from the lowest exciton state in atomically smooth
semiconductor quantum wires was first observed by Wegscheider et
al. \cite{W} in symmetric, T-shaped quantum wires made on the
intersection of two 70 {\AA} GaAs quantum wells surrounded by AlGaAs
with the Al fraction $x=0.35$. Recently the same group obtained
excitonic lasing in a 60 {\AA}/140 {\AA} asymmetric quantum wire with
a 7\% Al filled Stem well (see Figure \ref{asympot} \cite{R}).
They reported an interesting observation of two-mode lasing in this
structure. Under strong excitation they achieved simultaneous lasing
from two different states. There is a switching between those two
lasing modes as the temperature or pumping rate is changed. A simple
rate equation model (Chapter \ref{rate}) gives very good agreement
with experimental data, which suggests that we have lasing from two
different excitonic states in the structure.

All calculations published to date which include the Coulomb
interaction between the electron and hole have only examined the
ground state exciton. They have used either variational methods
\cite{Var1}--\cite{Var3} or other approximations \cite{Nonvar1,Nonvar2} 
and were performed only for symmetric wires and for very limited cases
realised experimentally in the early days of T-shaped wire
manufacturing. With the growing experimental realisation of these
structures as well as the interesting report of lasing phenomena there
is a need for accurate two-body calculations, treating on an equal
footing the single-particle potential and the Coulomb interaction, of
both the ground and excited states in the structure.

Excited states seem to be very important for the operation of
excitonic lasers \cite{R}. Calculations of energies, oscillator
strengths for radiative recombination (i.e, how the various states
couple to photons) as well as the full wave functions for the whole
spectra of interest would be very beneficial for understanding the
origins of certain transitions and effects. This could help in the
design of lasers with better properties and higher maximum
temperatures for excitonic lasing. The goal is to design excitonic
lasers which can operate at room temperature. Also, performing highly
accurate calculations of the ground state exciton in QWR and the
corresponding QW enables $E_b$ and $E_{con}$ to be obtained for
different geometries (both symmetric and asymmetric) for a wide range
of well widths and Al content, $x$. Such data are of great importance
for the optimisation of the structures.

In Chapter \ref{PRB} we present numerical calculations of
electron-hole states in a T-shaped quantum wire. Our method is based
on an exact numerical solution of the Schr\"{o}dinger equation in a
certain basis within the effective mass approximation. The method is
not restricted to a given number of excited states and we can
calculate as many of them as required. For some structures we have
calculated up to 100 excited states. We perform calculations for a
very wide range of T-shaped wires. In Section \ref{PRBmodel} the
numerical method is discussed in detail while in Section \ref{results}
we present the results. There we first study the spectra and wave
functions and present a discussion of the nature of the various
excited states. Finally we discuss $E_{con}$, $E_b$ and the difference
between the ground-state exciton energy and the first excited-state
energy, $E_{2-1}$, as a function of well width $Dx$ and Al molar
fraction $x$ for the symmetric and asymmetric quantum wires. Our
calculations are being used to design improved excitonic lasers which
will operate at room temperature.

In Chapter \ref{bell} we describe an experimental observation of
two-mode lasing in a 60 {\AA}/140 {\AA} asymmetric quantum wire with a
7\% Al filled Stem well \cite{R}. This experiment, done at
Bell-Laboratories, was a direct motivation for our work. We have
performed the numerical calculations described in Chapter \ref{PRB}
for samples used in this experiment and have identified the origin of
transitions which correspond to the two lasing modes.

In Chapter \ref{rate} we develop a rate equation model for a two-mode
laser in which the two lasing modes come from two different excitonic
states in the wire. This model gives very good agreement with the
experimental data described in Chapter \ref{bell}, which suggests that
we have lasing from two different excitonic states in the quantum
wire. The origin of this excitonic states can be determined from the
detail numerical calculations presented for this experiment in Chapter
\ref{bell}.

\chapter{Calculations of the electron-hole states in T-shaped quantum
wires}  
\chset{Calculations of the electron-hole states in T-shaped quantum
wires}
\label{PRB}

{ \it We calculate energies, oscillator strengths for radiative
recombination, and two-particle wave functions for the ground state
exciton and around 100 excited states in a T-shaped quantum wire. We
include the single-particle potential and the Coulomb interaction
between the electron and hole on an equal footing, and perform exact
diagonalisation of the two-particle problem within a finite basis
set. We calculate spectra for all of the experimentally studied cases
of T-shaped wires including symmetric and asymmetric
GaAs/Al$_{x}$Ga$_{1-x}$As and
In$_{y}$Ga$_{1-y}$As/Al$_{x}$Ga$_{1-x}$As structures. We study in
detail the shape of the wave functions to gain insight into the nature
of the various states for selected symmetric and asymmetric wires in
which laser emission has been experimentally observed. We also
calculate the binding energy of the ground state exciton and the
confinement energy of the 1D quantum-wire-exciton state with respect
to the 2D quantum-well exciton for a wide range of structures, varying
the well width and the Al molar fraction $x$. We find that the largest
binding energy of any wire constructed to date is 16.5 meV. We also
notice that in asymmetric structures, the confinement energy is
enhanced with respect to the symmetric forms with comparable
parameters but the binding energy of the exciton is then lower than in
the symmetric structures.  For GaAs/Al$_{x}$Ga$_{1-x}$As wires we
obtain an upper limit for the binding energy of around 25 meV in a 10
{\AA} wide GaAs/AlAs structure which suggests that other materials
must be explored in order to achieve room temperature applications.
There are some indications that
In$_{y}$Ga$_{1-y}$As/Al$_{x}$Ga$_{1-x}$As might be a good candidate. }

\section {The model}
\label{PRBmodel}

We use the effective mass approximation with an anisotropic hole mass
to describe an electron in a conduction band and a hole in a valence
band in the semiconductor structures under consideration (see Appendix
\ref{app2}). The effective mass of the hole depends on the
crystallographic direction in the plane of the T-shaped structure. We   
consider the heavy hole only. The other bands (split off bands, light
holes) would have energies higher that the region of interest for
us. The light hole exciton, the closest in energy to the heavy hole
exciton, is calculated to be over 30 meV higher that the heavy hole
exciton, and thus it is ignored in the calculations. The electron and
hole are in the external potential of the quantum wire formed at the
T-shaped intersection of the GaAs/Al$_{x}$Ga$_{1-x}$As quantum
wells. The so called Arm quantum well is grown in the 110 crystal
direction and intersects with a Stem quantum well grown in the 001
direction (see Figures \ref{sympot} and \ref{asympot}). In our model
the crystal directions 110, 001, and 110 correspond to $x$, $y$, and
$z$ respectively. We consider symmetric quantum wires where the Arm
and Stem well are both of the same width, i.e, $Dx=Dy$, and are made
of GaAs. We also consider asymmetric wires where the Stem well is
significantly wider but filled with Al$_{x}$Ga$_{1-x}$As with a low Al
content to compensate for the reduction in confinement energy. Our
method is applicable to any structure regardless of its shape and
materials provided the external potential is independent of $z$

The value of the band-gap is different for the different materials
used in the well construction. This gives rise to the potential
barriers at the interfaces between the GaAs, Al$_{x}$Ga$_{1-x}$As and
InGaAs which take different values for electrons and holes. In our
model the electron and hole are placed in external potentials
$V_e(x,y)$ and $V_h(x,y)$, respectively, and interact via the Coulomb
interaction. We choose the potential in GaAs to be zero and calculate
all potentials in other materials with respect to this level. The
external potential is independent of $z$ in all cases. Sample
geometries considered in this work are shown in Figs. \ref{sympot} and
\ref{asympot}. Using the above model, after the separation of the
centre of mass and relative motion in the $z$ direction, the system is
described by the following Hamiltonian:
\begin{multline}
 H =  
 -\frac{\hbar^2}{2m_e}\nabla^{2}_{x_{e},y_{e}}
 -\frac{\hbar^2}{2m_{hx}}\nabla^{2}_{x_{h}}
 -\frac{\hbar^2}{2m_{hy}}\nabla^{2}_{y_{h}}
 -\frac{\hbar^2}{2\mu_z}\nabla^{2}_{z}+ 
V_{e}(x_{e},y_{e}) + V_{h}(x_{h},y_{h})\\
 -\frac{e^2}{4 \pi \epsilon_0 \epsilon
 \sqrt{(x_{e}-x_{h})^{2}+(y_{e}-y_{h})^{2}+z^{2}}},
\end{multline}
where $z=z_e-z_h$ and
$\frac{1}{\mu_z}=\frac{1}{m_e}+\frac{1}{m_{hz}}$. The wave function
associated with the centre of mass motion in the $z$ direction is a
plane wave and this coordinate can be omitted from the problem.

\subsection {Numerical method for calculating quantum wire exciton states}

We calculate the ground and excited states in the structures of
interest by a direct diagonalisation method. Due to the complexity of
the external potential with its limited symmetry and sharp edges, none
of the standard basis sets seem appropriate. We use the following
basis set:
\begin{equation}
\label{basis-set}
\psi(x_e,y_e,x_h,y_h,z_e-z_h)=\sum_{i,j,k}c_{i,j,k}\sin(z\frac{k\pi}{L_z}-
\frac{k\pi}{2})\chi^e_i(x_e,y_e)\chi^h_j(x_h,y_h),
\end{equation}
where $\chi^e_i(x_e,y_e)$/$\chi^h_j(x_h,y_h)$ are electron/hole
single-particle wave functions for a T-shaped potential without the
electron-hole Coulomb interaction. In the $z$ direction we introduce
hard wall boundary conditions and use a standing-wave basis set.

Our basis set does not obey the so called cusp condition \cite{Kato}
which is satisfied whenever two particles come together. The
divergence in the potential energy when the electron and hole come
together must be exactly cancelled by an opposite divergence in the
kinetic energy. The exact wave function must therefore have a cusp
when the electron and hole are coincident. Using a basis in which
every basis function obeys the cusp condition would reduce the size of
the basis set required. For an isotropic hole mass it would be very
easy to satisfy the cusp condition by multiplying the basis functions
by the factor $e^{-\Lambda\sqrt{(x_e-x_h)^2+(y_e-y_h)^2+z^2}}$ which
is just the hydrogenic wave function. Unfortunately there is no
analytical solution when we introduce the anisotropic hole mass. Thus
we choose not to satisfy the cusp condition and therefore have to use
a larger basis set.

The diagonalisation is performed using a NAG library routine.
Convergence is usually achieved with a basis set containing 20 of each
of the single-particle wave functions and 20 standing waves in the $z$
direction. Thus $20\times20\times20$ = 8000 basis functions are needed
which gives $20^6$ matrix elements. Only one quarter of the total
number needs to be calculated as interchanging $k_1$ and $k_2$ leaves
the matrix element unchanged while interchanging $i_1$ and $j_1$ with
$i_2$ and $j_2$ gives its complex conjugate. This still leaves a great
many matrix elements to be calculated. Thus to make the calculations
feasible the matrix elements need to be calculated very rapidly (See
Section \ref{matrix}).

\subsection{Computational Method for Calculating the Single-Particle
Wave Functions}

\begin{figure}[htbp]
      \begin{center}
        \leavevmode
      \epsfxsize=14.0cm
      \epsfbox{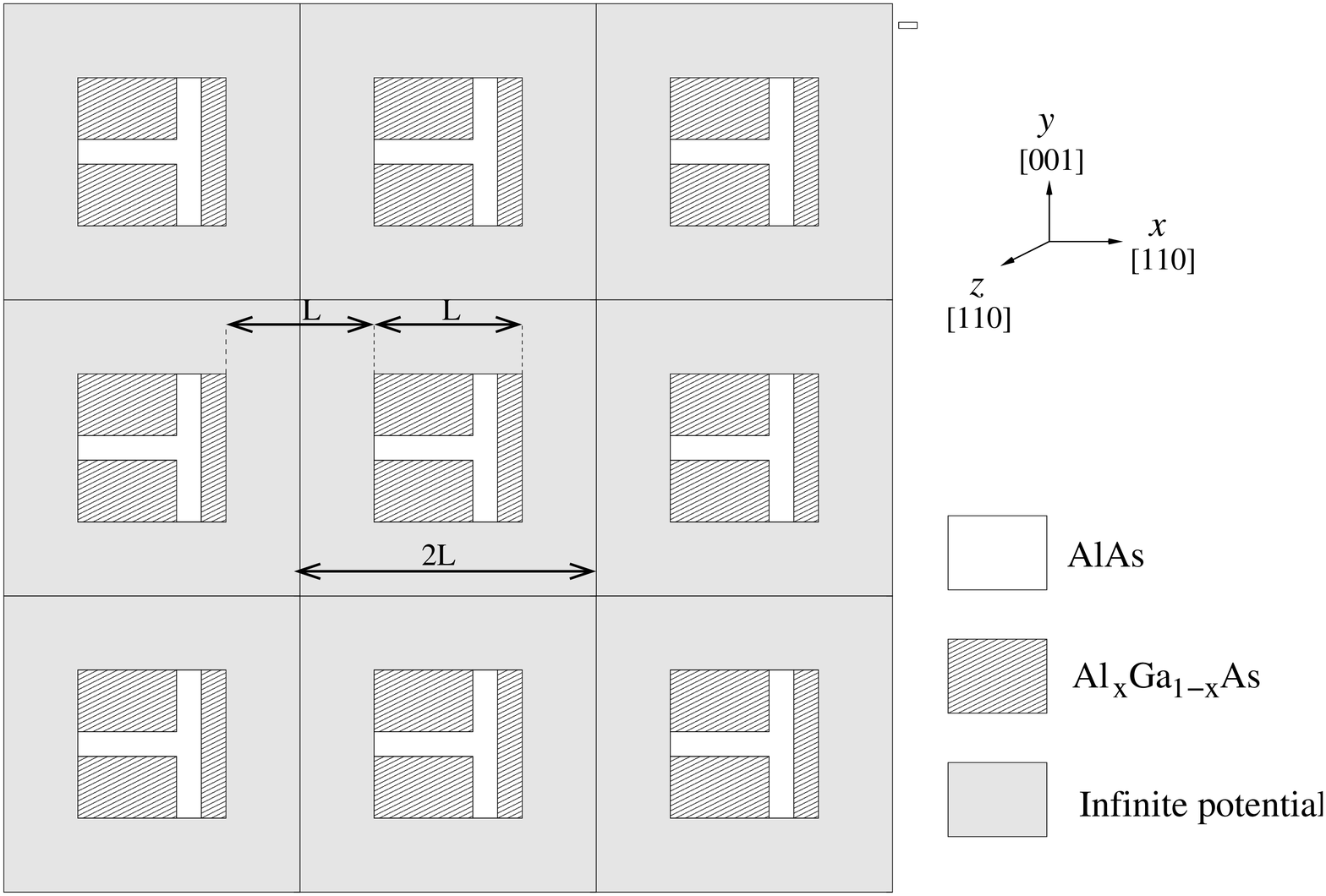}
        \end{center}
  \caption{Lattice used for calculations and notations. }
\label{lattice}
\end{figure}

The one-particle (electron and hole) wave functions,
$\chi^e_i(x_e,y_e)$ and $\chi^h_j(x_h,y_h)$ in a T-shaped external
potential are calculated using the conjugate-gradient minimisation
technique with pre-conditioning of the steepest descent vector. A
detailed explanation of this method can be found in reference
\cite{Payne}. We specify the external potential on a 2D grid and use
periodic boundary conditions in the $x$ and $y$ directions so that we
are able to use Fast Fourier Transform (FFT) methods to calculate the
kinetic energy in Fourier space while the potential energy matrix
elements are calculated in real space. The fast calculation of the
energy matrix elements is crucial as they have to be calculated many
times during the conjugate-gradient minimisation. The FFT provides
very fast switching between real and Fourier space and makes the
algorithm much more efficient, but the use of periodic boundary
conditions introduces the problem of inter-cell interactions in the
case of two particle calculations. To avoid this problem we place the
unit cell in the middle of another, larger unit cell of infinite
potential (see Figure \ref{lattice} and the Section \ref{matrix}).

We use plane waves as a basis set for the one-particle problem. Using
this method we can calculate as many as 50 states for the electron and
50 for the hole. Very good convergence with respect to the number of
plane waves and the size of the unit cell is obtained (see Section
\ref{accuracy}).

\subsection{Computational Method for Calculating the Matrix 
Elements}
\label{matrix}

The kinetic and potential energies are diagonal in this basis and are
obtained from the one-particle calculations. Thus only the Coulomb
matrix elements need to be calculated.

A Coulomb matrix element in the basis set (\ref{basis-set}) is a 5D
integral of the following form:
\begin{multline}
-\int\int\int\int\int dx_e dy_e dx_h dy_h dz  
sin(z\frac{k_2\pi}{L_z}- \frac{k_2\pi}{2})\chi^{e*}_{i_2}(x_e,y_e)
\chi^{h*}_{j_2}(x_h,y_h) \\ 
q(x_{e}-x_{h},y_{e}-y_{h},z)
sin(z\frac{k_1\pi}{L_z}- \frac{k_1\pi}{2})\chi^e_{i_1}(x_e,y_e)
\chi^h_{j_1}(x_h,y_h). 
\end{multline}
Where $q(x_{e}-x_{h},y_{e}-y_{h},z)$ is the Coulomb interaction cut
off at final distance to avoid image effects (see below). This
integral must be calculated numerically. Numerical integration for so
many dimensions is very slow and thus is not feasible for the case of
$20^6$ matrix elements. Thus another method has to be introduced.

The above integral is of the form 
\begin{equation}
-\int\int\int\int\int dx_e dy_e dx_h dy_h dz  
f_e(x_e,y_e)f_h(x_h,y_h)q(x_{e}-x_{h},y_{e}-y_{h},z)f_z(z).
\end{equation}
Where 
\begin{align}
f_e(x_e,y_e)&=\chi^{e*}_{i_2}(x_e,y_e)\chi^e_{i_1}(x_e,y_e),\nonumber\\
f_h(x_h,y_h)&=\chi^{h*}_{i_2}(x_h,y_h)\chi^h_{i_1}(x_h,y_h),\nonumber\\
\end{align}
Using the Fourier transform and the convolution theorem it can be
shown that the above integral is equal to:
\begin{equation}
\label{lastform}
\int dz \sum_{G_x,G_y}F_e(-G_x,-G_y)*F_h(G_x,G_y)*Q(G_x,G_y,z).
\end{equation}
Where $F_e$, $F_h$, $Q$ are the 2D Fourier transforms of the function
$f_e$ with respect to $x_e$ and $y_e$, $f_h$ with respect to $x_h$ and
$y_h$ and $q$ with respect to $x_e-x_h$ and $y_e-y_h$,
respectively. Thus the 5D integral can be reduced to a 1D integral
with respect to the $z$ variable and a 2D sum in Fourier space. The
$F_e$ and $F_h$ Fourier transforms can be easily calculated using FFTs
in real space after multiplication of the corresponding
$\chi^e_{i_1}(x_e,y_e)$ by $\chi^{e*}_{i_2}(x_e,y_e)$ for electrons
and $\chi^h_{i_1}(x_h,y_h)$ by $\chi^{h*}_{i_2}(x_h,y_h)$.
 
In order to use FFTs we need to introduce periodic boundary
conditions in the $x$ and $y$ directions as in the one-particle 
calculations. To eliminate interactions between particles in 
neighbouring cells, we place the unit cell in the middle of another,
bigger unit cell of infinite potential (see Figure \ref{lattice}).

The distance between the edges of successive small unit cells is
exactly the width of the small unit cell, $L$. We cut-off the Coulomb
interaction at a distance corresponding to the size of the small unit
cell. We therefor consider the following form of Coulomb interaction:
\begin{equation}
q(x_e-x_h,y_e-y_h,z)= \left\{ \begin{array}{ll}
-\frac{e^2}{4 \pi \epsilon_0 \epsilon
 \sqrt{(x_{e}-x_{h})^{2}+(y_{e}-y_{h})^{2}+z^{2}}}
& \parbox{3cm}{if $x_e-x_h < L_x$ and $y_e-y_h < L_y$} \\
0 & \mbox{otherwise}.
\end{array}
\right.
\end{equation}
Particles interact only when their separations in the $x$ and $y$
directions are smaller than $L_x$ and $L_y$ respectively. The
separations of particles in neighbouring cells is always bigger than
the cut-off and thus they do not interact. Particles in the same unit
cell are always separated by less than that the cut-off distance due
to the infinite potential outside the small unit cell. Thus we take
into account all of the physical Coulomb interaction and completely
eliminate the interactions between images. In the numerical
implementation the infinite potential is replaced by a large but
finite potential. Thus the probability of the particle being outside
the small unit cell is effectively zero and we find that the results
do not depend on the value of this potential for values greater than
around three times the potential in the Al$_{x}$Ga$_{1-x}$As region.

The 2D Fourier transform of the 3D Coulomb interaction with a cut-off
cannot be done analytically. Thus we put the Coulomb interaction onto
a 2D grid as a function of relative coordinates $x_{e}-x_{h}$ and
$y_{e}-y_{h}$ for every $z$ value. The unit cell in relative
coordinates will go from $-L_x$ to $L_x$, and $-L_y$ to $L_y$
respectively. Then for every value of $z$ a 2D FFT is performed with
respect to $x_{e}-x_{h}$ and $y_{e}-y_{h}$ and the results stored in
the 3D array $Q(G_x,G_y,z)$. Since this is the same for every matrix
element the above calculation needs to be performed only once.

The calculations described by eqn. (\ref{lastform}) need to be
performed for every matrix element. After $F_e(G_x,G_y)$ and
$F_h(G_x,G_y)$ have been calculated the summation over the reciprocal
lattice vectors $G_x$ and $G_y$ for every value of $z$ is
performed. The remaining 1D integral in the $z$ direction is done
numerically, after interpolation of data points, using a routine from
the NAG library. The dependence of the integrand on $z$ is found to be
very smooth and thus not many points are required to obtain accurate
results.

\section {Results}
\label{results}

We perform the calculations for a series of T-shaped structures.  We
calculate energies, oscillator strengths and wave functions for the
first 20-100 two-particle states for symmetric and asymmetric wires.

For symmetric wires we consider the structure denoted by W which has
been experimentally studied by Wegscheider et al. \cite{W} and
consists of GaAs/Al$_{0.35}$Ga$_{0.65}$As 70 {\AA} quantum
wells. Then, keeping the rest of parameters constant, we vary the
quantum well width from 10 {\AA} to 80 {\AA} in steps of 10 {\AA} in
order to examine the width dependence of the various properties. We
also perform calculations for samples denoted by S1 and S2 studied by
Someya et al. \cite{S1S2} made of GaAs/Al$_{0.3}$Ga$_{0.7}$As (S1) and
GaAs/AlAs (S2) quantum wells of width around 50 {\AA}. For the
GaAs/AlAs case we again vary the well width from 10 {\AA} to 60
{\AA}. Then we take an intermediate value of the Al molar fraction,
$x=0.56$, and vary the well width from 10 {\AA} to 60 {\AA} in order
to examine the dependence on the well width as well as Al
content. Finally we perform calculations for 35 {\AA}-scale
In$_{0.17}$Ga$_{0.83}$As/Al$_{0.3}$Ga$_{0.7}$As (denoted by N4) as
well as for 40 {\AA}-scale
In$_{0.09}$Ga$_{0.91}$As/Al$_{0.3}$Ga$_{0.7}$As (denoted by N2)
samples as studied experimentally by Akiyama et al. \cite{N2N4}.

For asymmetric structures we consider the wire studied experimentally
by Rubio et al. \cite{R} which consists of a 60 {\AA}
GaAs/Al$_{0.35}$Ga$_{0.65}$As Arm quantum well and a 140 {\AA}
Al$_{0.07}$Ga$_{0.93}$As/Al$_{0.35}$Ga$_{0.65}$As Stem quantum well. We
vary the width of the Arm quantum well from 50 {\AA} to 100 {\AA}. We
also perform calculations for the asymmetric structure studied by a
different group \cite{G1,G2} which consists of a 25 {\AA}
GaAs/Al$_{0.3}$Ga$_{0.7}$As Arm quantum well and a 120 {\AA}
Al$_{0.14}$Ga$_{0.86}$As/Al$_{0.3}$Ga$_{0.7}$As Stem quantum well.
  
In the first part of this section we present the spectra for symmetric
and asymmetric quantum wires with the positions of 2D exciton, 1D
continuum (unbound electron and hole both in the wire) and 1De/2Dh
continuum (unbound electron in the wire and hole in the well) states
as well as pictures of representative wave functions.  This allows us
to discuss the nature of the excited states in the structures. In the
second part we discuss the trends in confinement and binding energy
and the separation in energy between the ground and the first excited
states as a function of the well width and Al fraction.

We use a static dielectric constant $\epsilon$=13.2 and a conduction
band offset ratio $Q_c=\Delta E_{cond}/\Delta E_g$ of 0.65. For the
difference in bandgaps on the GaAs/Al$_x$Ga$_{1-x}$As interface we use
the following formula: $\Delta E_g=1247\times x$ meV for $x<0.45$ and
$1247 \times x+1147 \times (x-0.45)^2$ meV for $x>0.45$. For the
electron mass we use $m_e = 0.067 m_0$ while for the hole mass $m_{hx}
= m_{hz} = m_{h[110]} =0.69-0.71 m_0$ and $m_{hy}=m_{h[001]}=0.38 m_0$
($m_0$ is the electron rest mass). For the
In$_{0.09}$Ga$_{0.91}$As/Al$_{0.3}$Ga$_{0.7}$As
(In$_{0.17}$Ga$_{0.83}$As/Al$_{0.3}$Ga$_{0.7}$As) we use parameters
from reference \cite{N2N4}: for the electron $m_e=0.0647(0.0626)m_0$,
for the hole $m_{hy}=m_{hh[001]}=0.367(0.358)m_0$ and
$m_{hx}=m_{hz}=m_{h[110]}=0.682(0.656)m_0$, $\Delta E_g$=464(557) meV
and the band offset was assumed to be 65\% in the conduction and 35\%
in the valence band.

\subsection{Excited States}
\subsubsection{Symmetric Wires}
\label{symexstates}

\begin{figure}[p]
      \begin{center}
        \leavevmode
      \epsfxsize=14.0cm
      \epsfbox{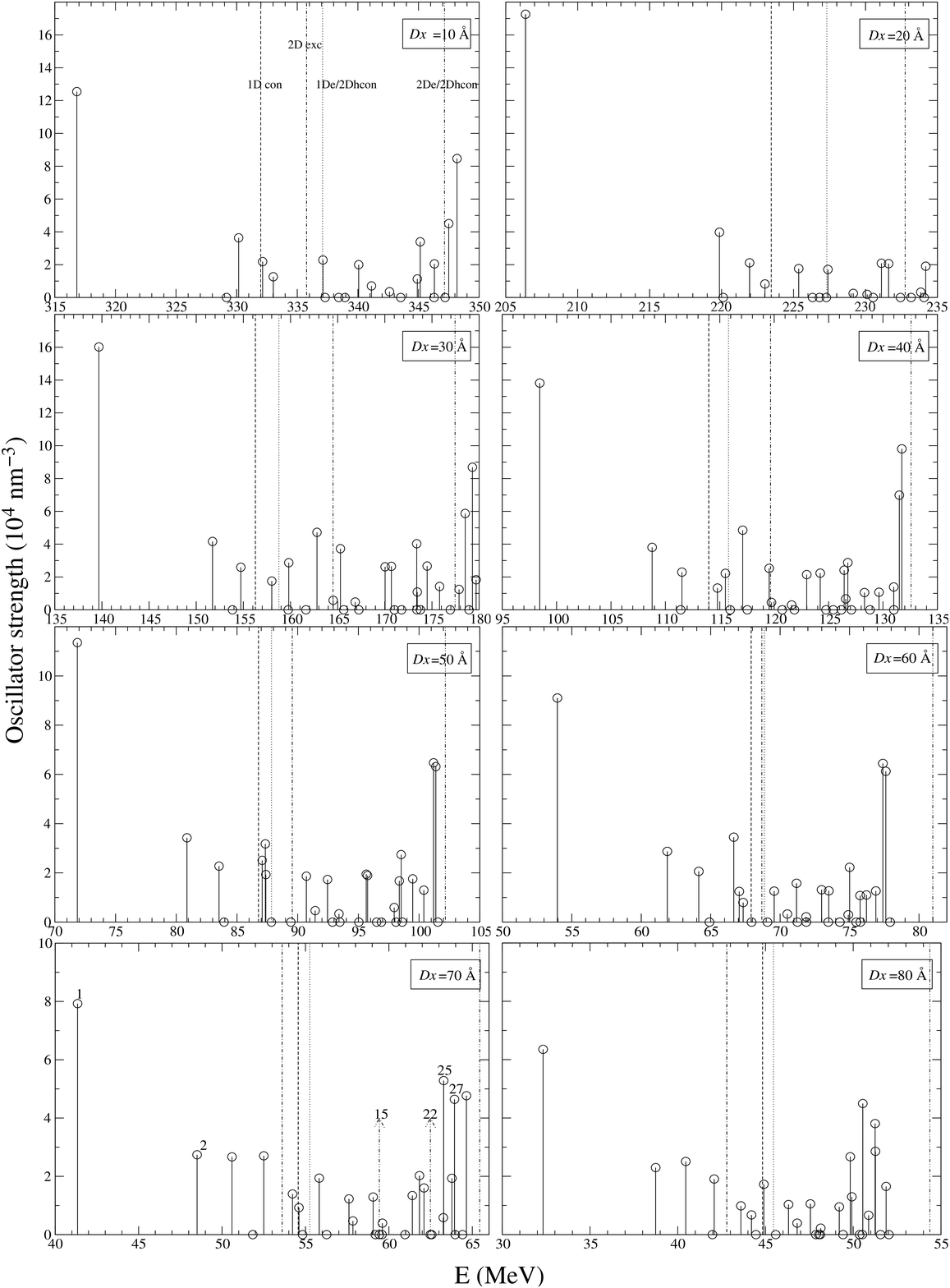}
        \end{center}
\caption{Oscillator strength versus energy for the lowest 20-30 states
in a symmetric T-shaped structure for different well widths $Dx$. }
\label{symspec}
\end{figure}

In Figure \ref{symspec} we show spectra (the oscillator strength
versus energy) for the first 20 (30 in the case of the 70 {\AA} and 30
{\AA} wire) states for the GaAs/Al$_{0.35}$Ga$_{0.65}$As structure for
well widths from 10 {\AA} to 80 {\AA}. 
A dashed line shows the energy of the 1D continuum, a dotted line that
of the 1D electron and 2D hole continuum, the dotted-dashed line - the
quantum-well 2D exciton, while the dashed-dot-dotted line shows the 2D
electron and 2D hole continuum.  In the case of the 20 {\AA} wire the
2D electron and 2D hole continuum is not shown as its value of 245.5
meV is out of range by a significant amount. Because our system is
finite in the $z$ direction, we obtain only a sampling of the
continuum states; below the continuum edge the states are discrete.

Note that for the experimentally studied 70 {\AA} structure, the 2D
exciton has a lower energy than the completely unbound electron and
hole in the wire. The situation clearly depends on the well width and
the crossing point is between 60 and 70 {\AA}. For well widths of 60
{\AA} or smaller, the 1D continuum (1Dcon) is lower in energy that the
2D exciton (2Dexc) with the difference being maximal for a width of
around 20 {\AA}. For widths of 70 {\AA} or bigger, the 1Dcon is higher
in energy than the 2Dexc with the difference growing for increasing
well width. This effect might be significant for pumping T-shaped-wire
lasers. Free electrons and holes are excited in the whole area of both
wells and thus, when the 2D exciton has a lower energy than the 1D
continuum, formation of the 2D excitons is energetically favourable.
These excitons can recombine in a well instead of going to the wire
and forming a 1D exciton. Clearly it is more efficient to have the 1D
continuum lower in energy than the 2D exciton.

By increasing the well width we obtain more states that are lower in
energy than the 1Dcon and 2Dexc beginning with two (ground and the
first excited) for the 10 {\AA} well, three for widths between 20-50
{\AA} and four states for larger widths.

\begin{figure}[p]
      \begin{center}
        \leavevmode
      \epsfxsize=14.0cm
      \epsfbox{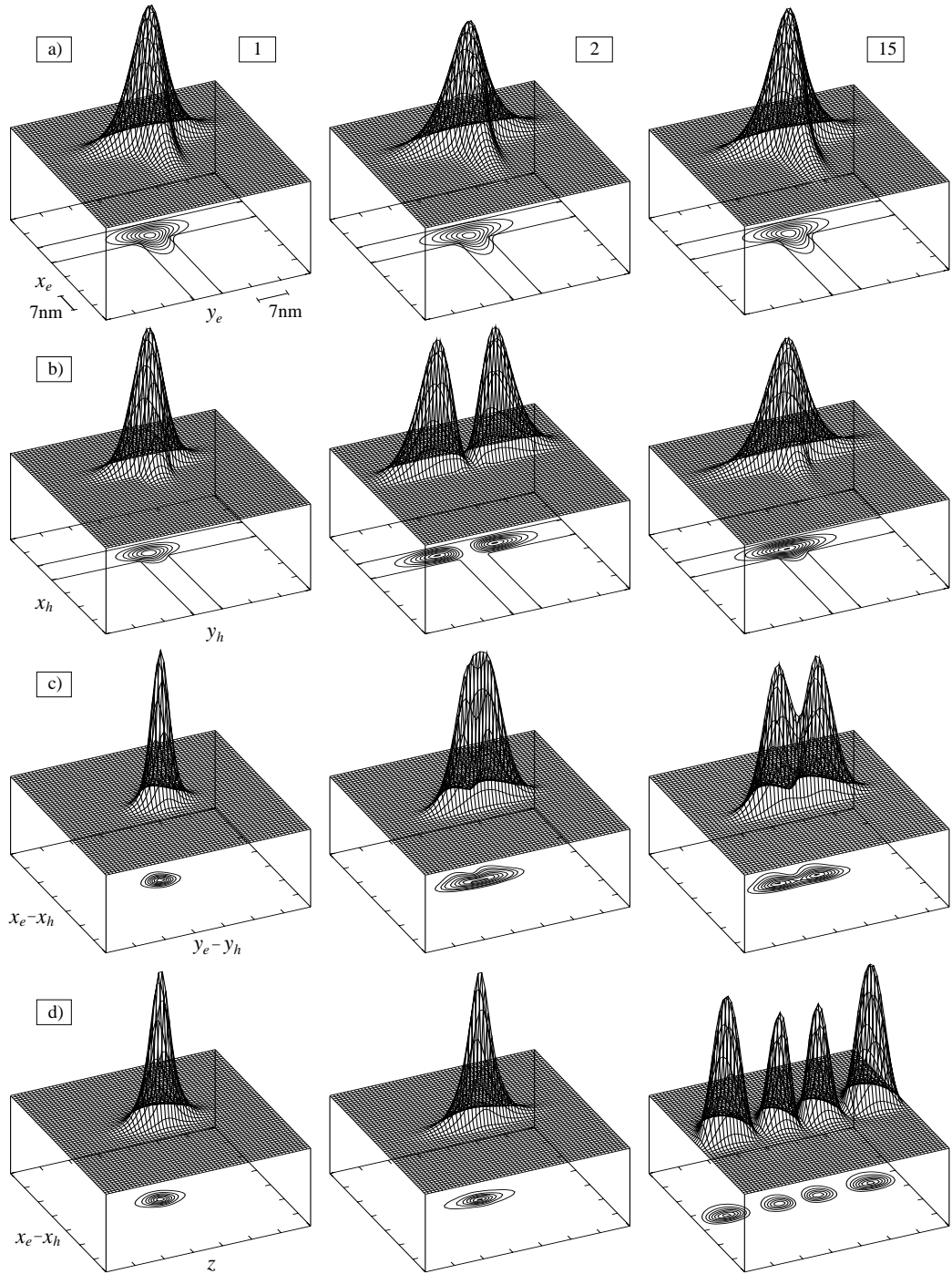}   
        \end{center}
\caption{Modulus squared of the two-particle wave function for the
ground (1), first excited (2) and the 15th (15) state in the symmetric
T-structure. Electron (a), hole (b) and the relative coordinates
$x_e-x_h$, $y_e-y_h$ (c), $x_e-x_h$, $z$ (d) probability densities are
shown. }
\label{sym70st1}
\end{figure}

\begin{figure}[p]
      \begin{center}
        \leavevmode
      \epsfxsize=14.0cm
      \epsfbox{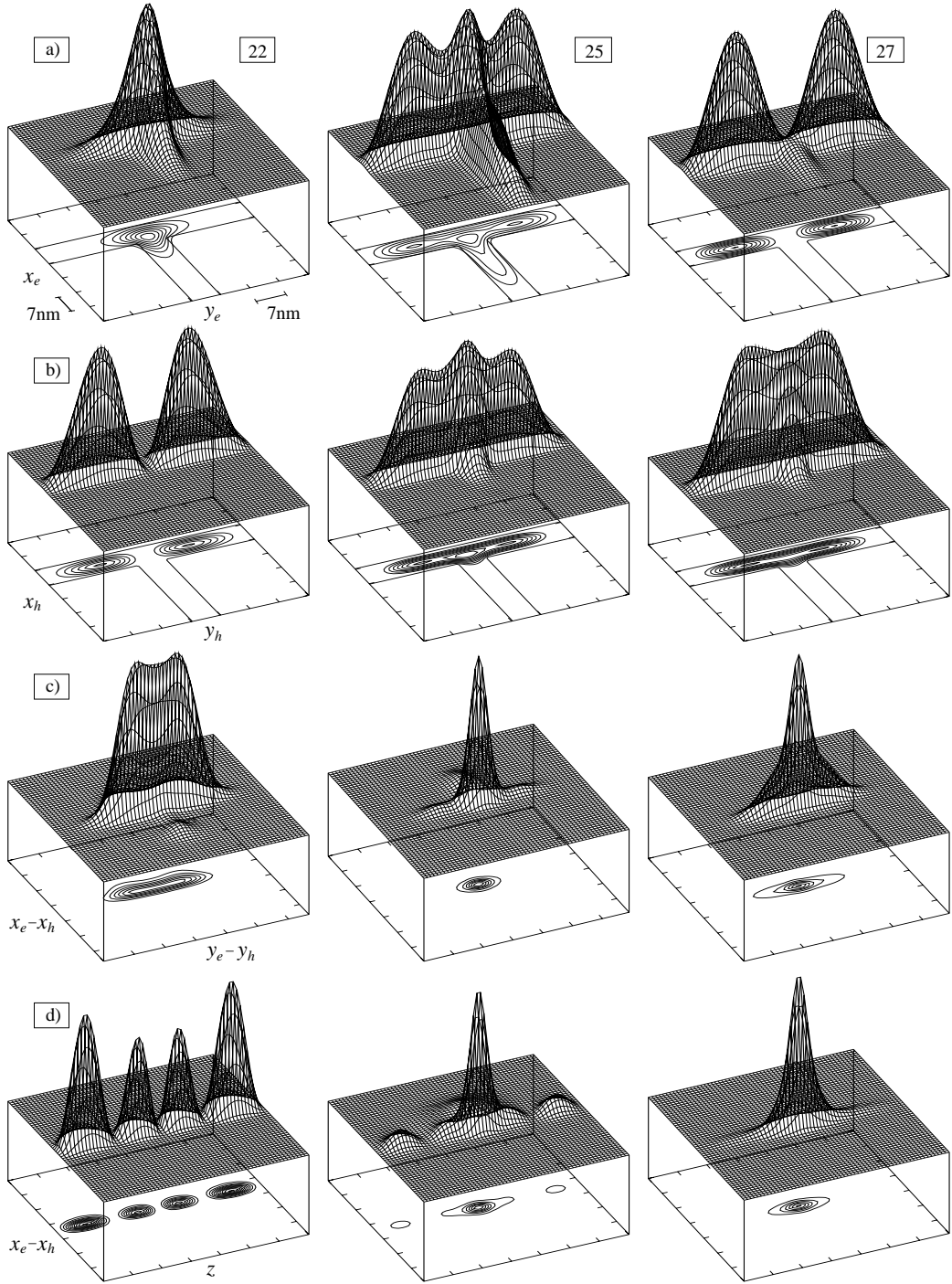}   
        \end{center}
\caption{Modulus squared of the two-particle wave function for the
22th, 25th and the 27th state in the symmetric T-structure. Electron
(a), hole (b) and the relative coordinates $x_e-x_h$, $y_e-y_h$ (c),
$x_e-x_h$, $z$ (d) probability densities are shown.}
\label{sym70st2}
\end{figure}

We now discuss the behaviour of $|\psi|^2$ for the 70 {\AA} case. The
wave functions depend on five spatial coordinates and thus various
cuts in 5D space are presented in Figures \ref{sym70st1} and
\ref{sym70st2}: a) the electron $x_e$, $y_e$ position after averaging
over the hole position, b) the hole $x_h$, $y_h$ position after
averaging over the electron position, and relative coordinates after
averaging over the centre of mass position c) the $x_e-x_h$, $y_e-y_h$
relative coordinates for $z_e-z_h=0$ and d) the $x_e-x_h$, $z$ relative
coordinates for $y_e-y_h=0$.

For the ground state we observe that the electron and hole are very
well localised in the wire with slightly more hole localisation. The
relative coordinate plots clearly show the bound exciton (Figure
\ref{sym70st1}(1)).

The electron in the first excited state is localised in the wire while
the hole already expands into the Arm well. The relative coordinate
pictures show that the electron and hole are bound and form an exciton
with an asymmetric shape. The size of the exciton is smallest in the
$x$ direction (the Stem well direction) and the exciton expands more
into the $y$ (the Arm well where the hole is expanded) and free $z$
directions (Figure \ref{sym70st1} (2)). The oscillator strength of
this state is about 1/3 of that of the ground state and the state
clearly takes the form of a 1D exciton with its centre of mass in the
T-wire.

It can be seen from the spectra (Figure \ref{symspec}) that there are
four states (apart from the ground state) with energies smaller than
1Dcon and 2Dexc. The nature of the 3rd and 5th states is very similar
to the 2nd one: the centre of mass is in the wire and the electron is
still well localised in the wire while the hole spreads into the wells
(into both the Arm and Stem wells for the 3rd state while only into
the Stem well for the 5th one). The relative coordinates show the
complex, asymmetric shape of this excitonic state and the oscillator
strength is again around 1/3 of the ground state exciton.

The 4th state with almost zero oscillator strength corresponds to a 1D
continuum. The electron and hole are both in the wire but the relative
coordinate pictures show an unbound exciton. Within the first 30
states we have 3 states of that nature: the 4th, 7th and 15th. The
15th state is shown in Figure \ref{sym70st1}: the electron and hole
are confined in the wire (a, b) and there are 3 nodes in the $z$
direction and 1 node in the $y$ direction. The other two states look
similar and differ only in the number of nodes. The energy of the 4th
state, which is the lowest 1Dcon state, turns out to be lower than the
real 1Dcon obtained from our one-particle calculations. This is due to
the finite size effects. Our method is very well converged with
respect to the cell size for the bound state and for the unbound ones
where at least one of the particles is in the well. However, for the
unbound continuum 1D states, the particles are very close in the $x,y$
plane because of the very small size of the wire and thus the
interaction is stronger. Consequently it does not decay as fast in
the $z$ direction as other states and thus we need a much bigger unit
cell in the $z$ direction to achieve convergence. There are however
only three such states within the 30 we examine and we know their true
energies from the preceding one particle calculations.

For further excited states up to the 25th, the electron, and thus the
centre of mass, is still localised in the wire while the hole is
taking up more and more energetic states in both wells, where energies
are quantised due to the finite size of the cell. Those states can be
divided into two groups depending on their relative coordinate nature:
excitonic-like states similar to the second state (Figure
\ref{sym70st1} (2)) and ionised states like the 22nd which is
represented in Figure \ref{sym70st2} (22). The oscillator strength of
the second group is zero (see Figure \ref{symspec}).

The 25th state (Figure \ref{sym70st2} (25)) is the first state where the
electron is de-localised in both wells, the relative coordinates
and the large oscillator strength shows that it is clearly an
excitonic-like state. It appears to be a 2D quantum well exciton state
scattered on the T shaped intersection. Its energy is thus higher than
that of a pure 2Dexc.

The 27th state is the 2D Arm-quantum-well-exciton state. It has higher
energy than the ground-state 2D exciton because the  electron and hole
wave functions occupy higher energy states than the ground state of the
well due to the presence of the T intersection.

The 30th state has a very similar nature to the 27th but the exciton
expands into the Stem instead of the Arm quantum well.

The 25th, 27th and 30th states all have large oscillator strengths
(around 3/4 of that of the ground state exciton). It is interesting to
note that between the ground state and those 2D
large-oscillator-strength states, there is a group of states with
relatively low oscillator strengths. The reason for this is that after
the ground state, there are states where either the wire-like electron
is bound to the well-like hole and thus they do not overlap enough to
give big contribution to the spectrum or they consist of a wire like
electron with an unbound hole.


Those quantum-well-like exciton states that scattered on the T-shaped
potential (like state 25) appear to be quite important for the
excitonic lasing because of their big oscillator strength. In \cite{R}
the authors reported two-mode lasing in an asymmetric wire where the laser
switches between the ground-state exciton and the other state whose
energy corresponds to the state from the tail of the above mentioned
states.

\subsubsection{Asymmetric Wires}

\begin{figure}[htbp]
      \begin{center}
        \leavevmode
      \epsfxsize=13.0cm
      \epsfbox{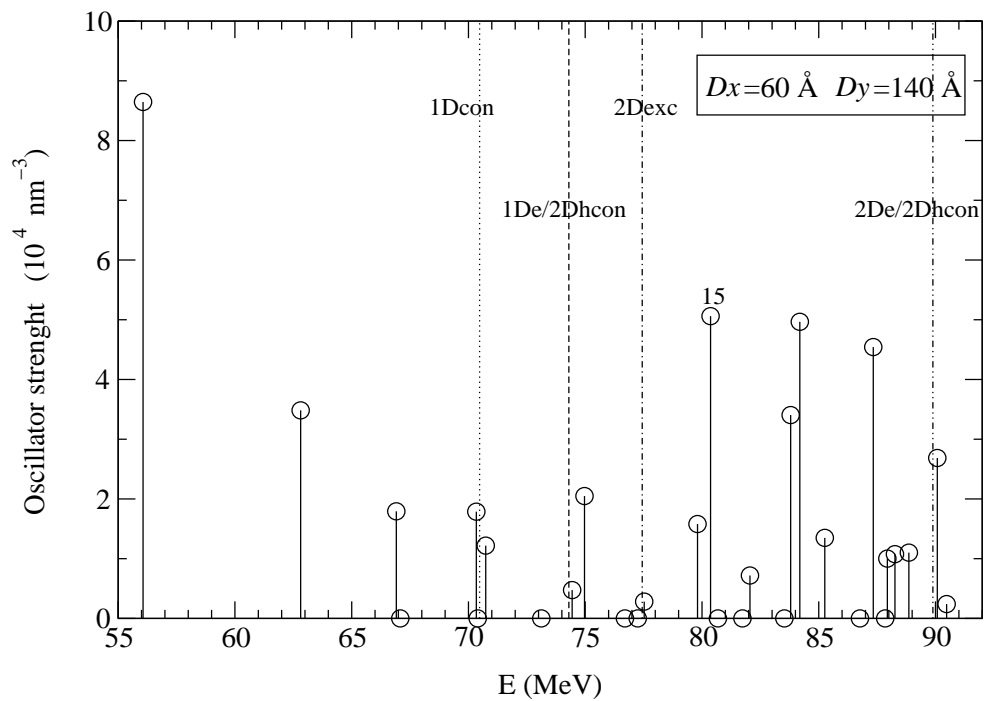}   
        \end{center}
\caption{Oscillator strength versus energy for the lowest 30 states in an
asymmetric T-shaped structure with $Dx$ = 60 {\AA}, $Dy$ = 140 {\AA}.}
\label{asymspec}
\end{figure}

The asymmetric wire that we study in detail consists of a 60 {\AA} or
56 {\AA} GaAs/Al$_{0.35}$Ga$_{0.65}$As Arm quantum well and a 140 {\AA}
Al$_{0.07}$Ga$_{0.93}$As/Al$_{0.35}$Ga$_{0.65}$As Stem quantum well
\cite{R}. The spectrum for the 60 {\AA} Arm case is shown in Figure
\ref{asymspec}. 
The nature of the states is very similar to the case of the symmetric
wire. The first two excited states are exciton-like and have an
electron confined in the wire while the hole spreads into the
well. All excited states up to the 20th have the electron confined in
the wire. The hole spreads to one or both quantum wells taking up more
energetic states in the well. The relative coordinates show either an
exciton-like wave function (states with nonzero oscillator strength in
the spectra of Figure \ref{asymspec}) or the case where a hole is
confined in the wire but is not bound to the electron (states with
zero oscillator strength in the spectra). Both groups were discussed
and shown for the symmetric wire.

The 21st and the 24th states (large oscillator strengths in the Figure
\ref{asymspec}) have an electron expanding into the Arm well. The
electron wave function has a node in the wire region. The hole
wave function spreads into the Arm well and has no node for the 21st
state and one node in the wire region for the 24th state. The relative
coordinates show the excitonic nature of these states.  Thus these
states correspond to those 2D excitonic states scattered on the wire.

\begin{figure}[p]
      \begin{center}
        \leavevmode
      \epsfxsize=7.0cm
      \epsfbox{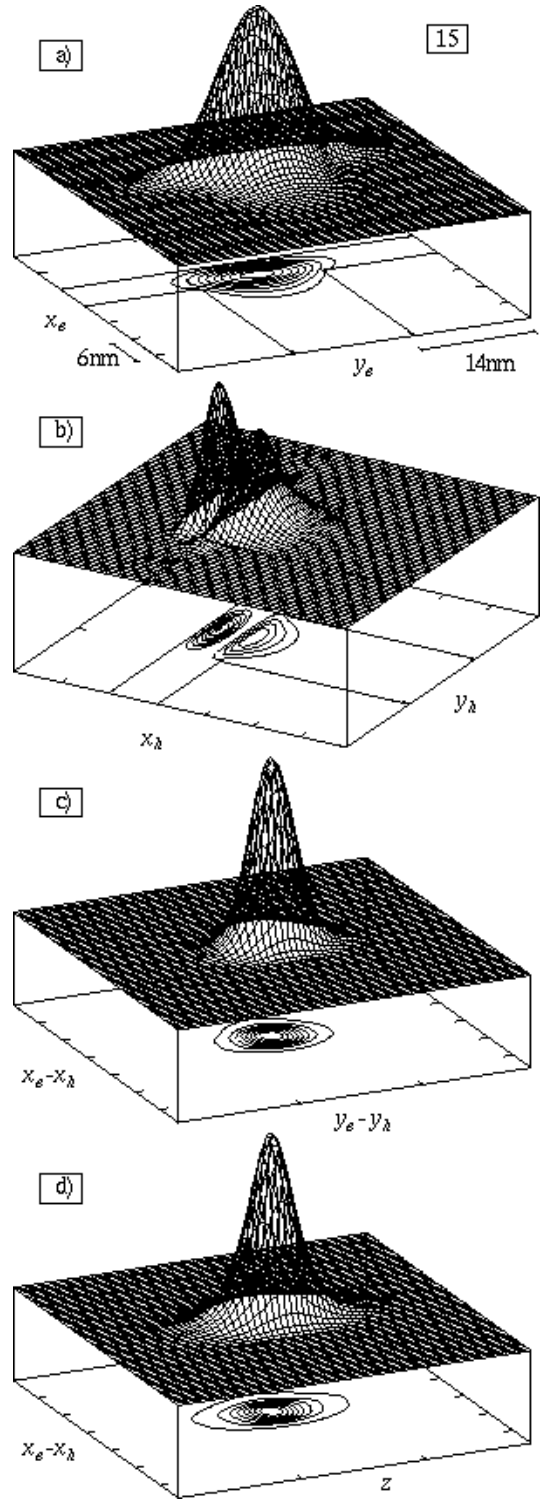}
        \end{center}
\caption{Electron (a), hole (b), and the relative coordinates
$x_e-x_h$, $y_e-y_h$ (c), $x_e-x_h$, $z$ (d) probability densities for
the 15th state in an asymmetric T-shaped structure with $Dx$ = 60
{\AA}, $Dy$ = 140 {\AA}.}
\label{asym15}
\end{figure}

For the asymmetric structure we observe one state (the 15th, see
Figure \ref{asym15}) which does not correspond to any state in the
symmetric case. The state is clearly excitonic-like with a large
oscillator strength and the relative coordinate plots show a very well
bound exciton. The electron is confined in the wire in the same way as
the ground state while the hole is clearly 1D-like, strongly confined
in the wire but in a different way. It has a node in the wire region.

\subsection{Trends in Confinement and Binding Energies}
\label{trends}

\subsubsection{Symmetric Wires}

\begin{figure}[p]
      \begin{center}  
        \leavevmode
      \epsfxsize=9.5cm
      \epsfbox{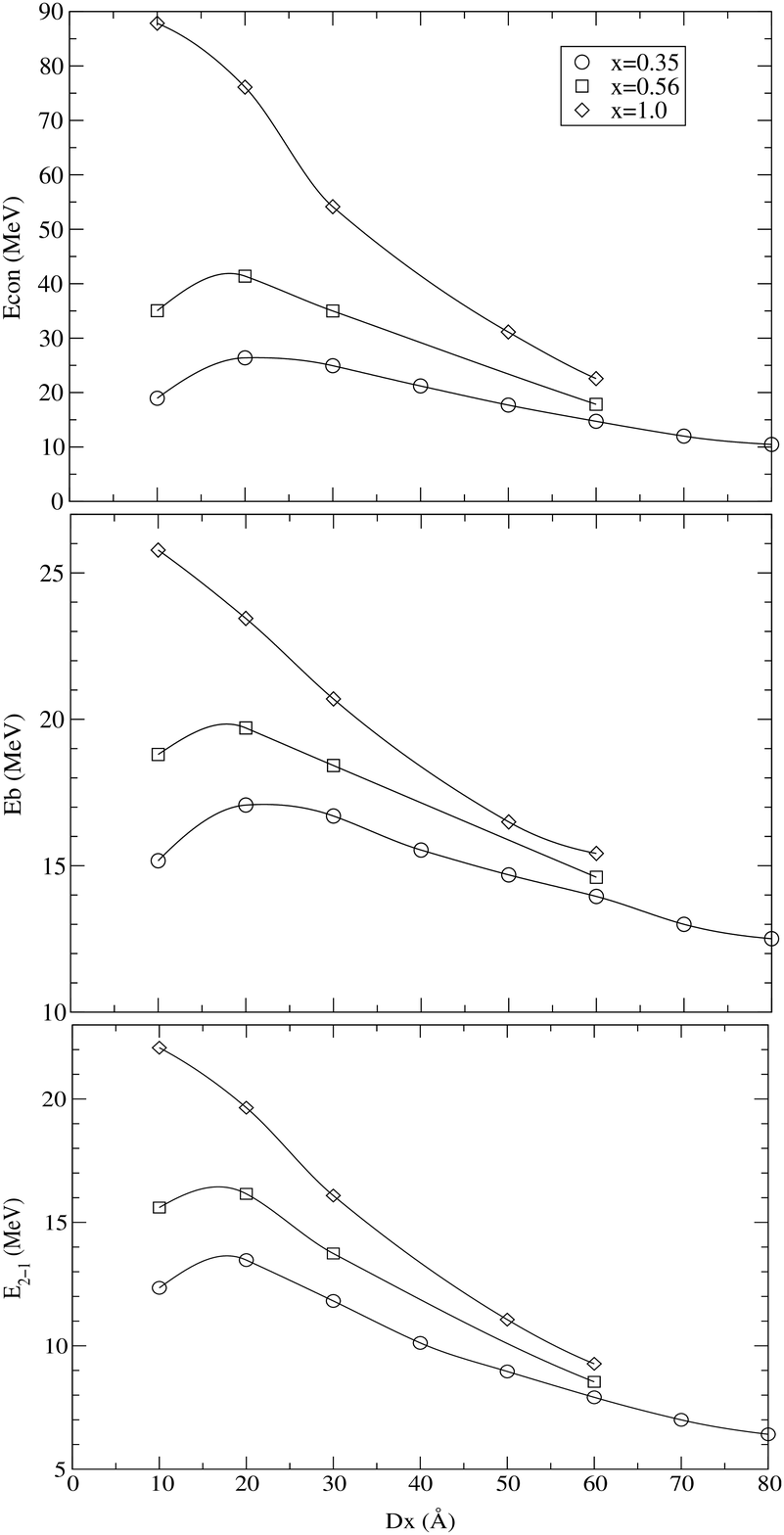}
        \end{center}
\caption{Confinement energy $E_{con}=E_{2Dexc}-E_{1Dexc}$, binding
energy of the ground-state exciton $E_b$, and the energy difference
between the ground state and the first excited state $E_{2-1}$ as a
function of the well width $Dx$ in a symmetric T-structure for three
different aluminium molar fractions $x$. }
\label{symEcEb}
\end{figure}

We calculate the exciton binding energy, $E_b=E_e+E_h-E_{1Dexc}$,
where $E_e$ and $E_h$ are the one-particle energies of an electron and
a hole, respectively, in the wire. We also calculate, using the same
method, the exciton energy in the quantum well, $E_{2Dexc}$, to obtain
the confinement energy of the 1D exciton, $E_{con}=E_{2Dexc}-E_{1Dexc}$,
in the wire.
We perform calculations for a wide range of structural parameters. For
the GaAs/Al$_{x}$Ga$_{1-x}$As quantum wire we change the well width
from 10 {\AA} to 80 {\AA} for three different values of the Al content
$x$. The results are shown in Figure \ref{symEcEb}. 
It can be noticed that for a well width bigger than 50 {\AA}, changing
the Al content has very little effect on the confinement and binding
energies. The difference in binding energy between the 60 {\AA}
GaAs/Al$_{0.35}$Ga$_{0.65}$As and the pure AlAs is only 1.5 meV. Thus
it seems more promising to change the well width rather than the Al
content for relatively wide wires. However, for thinner wires in the
range of 10-50 {\AA}, changing the Al content is much more profitable
then changing the well width. The difference in binding energies for
20 {\AA} wires with Al molar fractions of $x$=0.3 and $x$=1.0 is 6.4
meV. This increases to 10.6 meV when the width is reduced to 10 {\AA}.

$E_b$ and $E_{con}$ for Al contents of $x$=0.35 and $x$=0.56 both
approach a maximum for a well width between 10 {\AA} and 20 {\AA}. The
maximum values for $x$=0.35 are $E_{bmax}=17.1$ meV, $Econ_{max}=26.4$
meV and for $x$=0.56 they are $E_{bmax}=19.7$ meV, $Econ_{max}=41.4$
meV. For the $x$=1.0 case, the curve does not have a maximum in the
region for which calculations has been performed but we consider going
to wells thinner than 10 {\AA} as practically uninteresting. Thus the
maximum energies are for $Dx$=10 {\AA} and they are $E_{bmax}=25.8$ meV and
$Econ_{max}=87.8$ meV. 

$E_{con}$ increases much more rapidly than $E_b$ when the well width
is progressively reduced. The curves cross for a well width between 60
{\AA} and 70 {\AA}, i.e. for widths of 60 {\AA} or smaller, $E_{con}$
is greater that $E_b$ which means that the 1D continuum is lower in
energy than the 2D exciton (as we discussed in section
\ref{symexstates}) with the difference having a maximum at around 20
{\AA}. For widths of 70 {\AA} or bigger, $E_b$ is greater than
$E_{con}$ with the difference growing for increasing well width. We
also consider the difference in energy between the ground state
exciton in the wire and the first excited state as a function of the
well widths. For the experimentally realised $Dx$ = 70 {\AA} case,
this difference is $E_{2-1}=7.0$ meV and the maximum value for $Dx$ =
10 {\AA} is $E_{2-1max}=13.5$ meV. The maximum value for the GaAs/AlAs
at $Dx$ = 20 {\AA} is 22 meV.

Although pure AlAs gives the biggest potential offsets and thus the
biggest binding and confinement energies, the GaAs/AlAs interfaces are
not very smooth, which influences the transport properties. Thus new
materials have to be proposed. Two structures based on InGaAs have
been manufactured and measured \cite{N2N4}: 35 {\AA}-scale
In$_{0.17}$Ga$_{0.83}$As/Al$_{0.3}$Ga$_{0.7}$As (N4) and 40 {\AA}-scale
In$_{0.09}$Ga$_{0.91}$As/Al$_{0.3}$Ga$_{0.7}$As (N2). The results of
calculations for these structures are presented in Table \ref{table}.
It can be seen that energies for the sample N4 are almost exactly
the same as for the GaAs/AlAs sample S2 suggesting that these
materials might be very good candidates for structures with large exciton
confinement and binding energies.

\subsubsection{Asymmetric Wires}

In order to increase binding and confinement energies, the asymmetric
T-shaped structure was proposed and realised by two groups
\cite{R,G1,G2}. 

\begin{figure}[htbp]
      \begin{center}  
        \leavevmode 
      \epsfxsize=14.0cm 
      \epsfbox{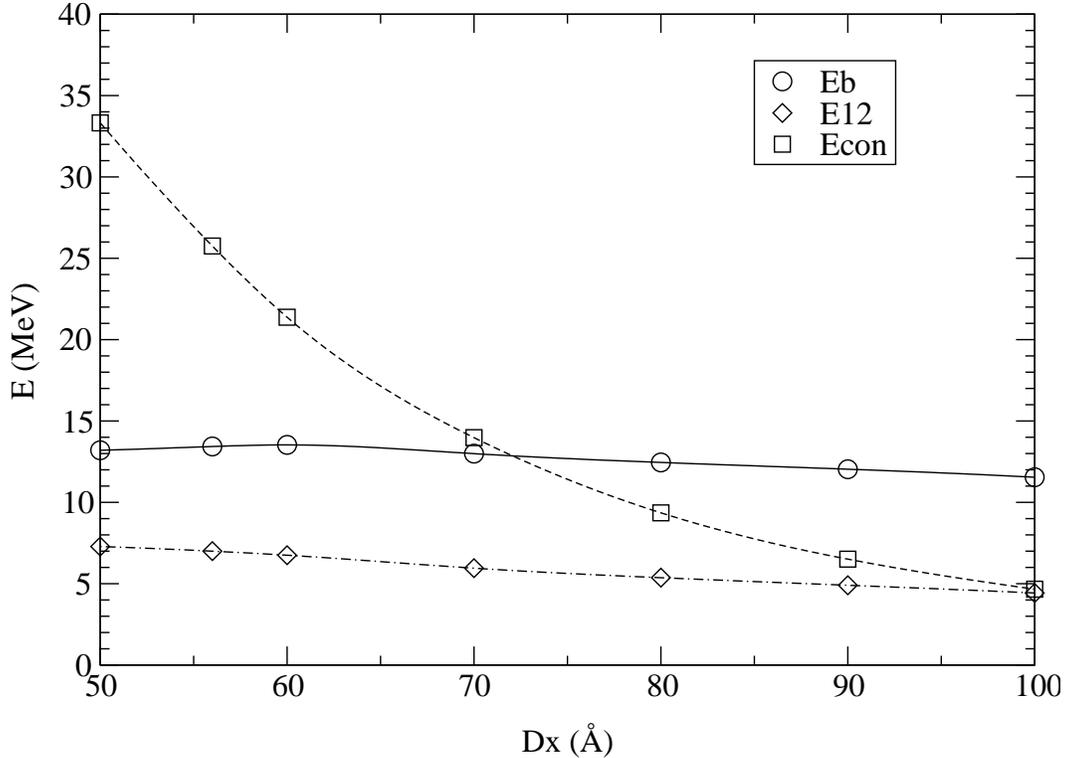} 
        \end{center} 
\caption{Confinement energy $E_{con}=E_{2Dexc}-E_{1Dexc}$, binding
energy of the ground state exciton $E_b$, and the difference between the
ground and the first excited state $E_{2-1}$ for an asymmetric
wire as a function of the well width $Dx$, where $Dy$ = 140 {\AA}.}
\label{nasEbEc}
\end{figure}

We calculate $E_b$ and $E_{con}$ for the 60 {\AA}/140 {\AA} structure
with the Stem quantum well filled with 7\% Al in order to compare with
experiment \cite{R} and then we vary the width of the Arm well from 50
to 100 {\AA}. One can see from Figure \ref{nasEbEc} that the binding
energy is almost independent of the Arm well width in this region,
changing only from the maximum value of 13.5 meV for $Dx$ = 60 {\AA}
to 11.5 meV for $Dx$ = 100 {\AA}. 
The binding energy for the 60 {\AA} symmetric wire with the same
$x$=0.35 Al mole fraction is 13.9 meV - a bit bigger than for the
asymmetric structure. In contrast, the confinement energy, $E_{con}$,
changes rapidly with the width of the Arm well from 4.7 meV for $Dx$ =
100 {\AA} up to 33.3 meV for $Dx$ = 50 {\AA}. For Arm-well widths of
60 {\AA} or bigger, the 2D quantum-well exciton in the Arm well has a
lower energy than that for the Stem well, thus the confinement energy
is calculated with respect to the Arm well exciton. For the 50
{\AA}-wide Arm well, the 2D exciton has higher energy than for the
Stem quantum well and thus the confinement energy is calculated with
respect to the Stem quantum well. Therefore 33.3 meV is the highest
confinement energy for this Stem well and changing the Arm well would
have no effect. Thus the 60 {\AA}/140 {\AA} structure is well
optimised and its confinement energy, $E_{con}$, is 21.4 meV which is
much bigger than that of 14.7 meV for the 60 {\AA} symmetric wire.

The highest confinement energy so far reported is for an asymmetric
GaAs/Al$_{0.35}$Ga$_{0.65}$As wire with a 25 {\AA} Arm quantum well and
a 120 {\AA} Stem quantum well filled with 14\% Al \cite{G1,G2}. The
experimentally obtained $E_{con}$ for this structure is 54 meV.  Our
calculations however give only 36.4 meV which is still the highest
among experimentally obtained structures but much lower than 
reported by the authors. Our calculation of $E_{con}$ for five different
experimentally realised structures agree very well with the
experimental values and thus it is very probable that the value of 54 
meV is overestimated. The binding energy from our calculations is only
14.6 meV for this structure.

We can conclude from our results that the optimised asymmetric
structure does not lead to a bigger exciton binding energy than the
symmetric ones with the same parameters. The confinement energy is
considerably enhanced and this effect, which can be measured directly,
has often been used to infer that the binding energy is increased.
However, our results show that no such relationship holds between the
confinement and binding energies.  Thus the biggest confinement energy
of any structure constructed so far of 36.4 meV does not lead to the
biggest binding energy. Indeed, the binding energy of 14.6 meV is
smaller than the 16.5 meV reported for the GaAs/AlAs 50 {\AA}-scale
symmetric structure \cite{S1S2} where the confinement energy should be
only 31.1 meV. It is also smaller than expected for a symmetric 25
{\AA}-scale structure with the same parameters (16.0-16.5 meV). Thus
asymmetric structures could be useful for applications where a large
confinement energy is required but appear to be less suitable than
symmetric wires for applications where large binding energies are of
interest.

\subsubsection{Comparison with Experiment and Other Calculations.}

The comparison between experiment and other published calculations is
presented in Table \ref{table}. The confinement energy of the exciton
can be directly measured experimentally. Although, due to the strong
inhomogeneous broadening of the photoluminescence peaks, the accuracy
of this number is not very high, it is the only experimentally proven
quantity we can refer to. The experimental binding energy needs to be
calculated using both experimental data and one-particle calculations
and thus errors might accumulate. Other theoretical methods which we
refer to obtain the ground state exciton energy using variational
techniques \cite{Var1}-\cite{Var3} (they differ in the form used for
the variational wave functions). There are also two non-variational
calculations for the ground state exciton \cite{Nonvar1,Nonvar2}.

\begin{landscape}
\begin{table}[p]
\caption{Binding energy, $E_b$ and the confinement
energy $E_{con}=E_{1Dexc}-E_{2Dexc}$ in meV of the QWR exciton for
seven different samples $W, S_1, S_2, N_2, N_4, R$ and $G$ obtained from
different methods.}
\begin{center}
\begin{tabular}{lcccccccccccccc}
\multicolumn{15}{l}{} \\
\hline
\hline
\multicolumn{15}{l}{} \\
&\multicolumn{2}{c}{W$^\mathrm{a}$
}
&\multicolumn{2}{c}{$S_1$$^\mathrm{b}$
}
&\multicolumn{2}{c}{$S_2$$^\mathrm{b}$
}
&\multicolumn{2}{c}{$N_2$$^\mathrm{c}$
}
&\multicolumn{2}{c}{$N_4$$^\mathrm{c}$
}
&\multicolumn{2}{c}{$R$$^\mathrm{d}$
}
&\multicolumn{2}{c}{$G$$^\mathrm{e}$
}\\
Method&$E_b$&$E_{con}$&$E_b$&$E_{con}$&$E_b$&$E_{con}$&$E_b$&$E_{con}$&$E_b$
&$E_{con}$&$E_b$&$E_{con}$&$E_b$&$E_{con}$\\
\hline
\multicolumn{15}{l}{} \\
Exp$^\mathrm{f}$
&17&17&17&18&27&38&&28&&34&13.8&23&&54 \\
 This work&13&12&14.3&17.8&16.5&31.1&12.1&26.3&16.5&31.2&13.5&21.4&14.6&36.4\\
 Nonvar1$^\mathrm{g}$
&13.2&&14.3&&16.4\\ 
 Nonvar2$^\mathrm{h}$
&&&11.63&&13.9\\
 Var1$^\mathrm{i}$
&&&15&&18\\
 Var2$^\mathrm{j}$
&9.6&11.9\\
Var3$^\mathrm{k}$
&12&14\\
\hline
\hline
\multicolumn{15}{l}{} \\
\multicolumn{15}{l}{\small $^\mathrm{a}$ Sample and experimental
values from Ref.\cite{W}.} \\ 
\multicolumn{15}{l}{\small $^\mathrm{b}$ Sample and experimental
values from Ref.\cite{S1S2}.} \\  
\multicolumn{15}{l}{\small $^\mathrm{c}$ Sample and experimental
values from Ref.\cite{N2N4}. } \\ 
\multicolumn{15}{l}{\small $^\mathrm{d}$ Sample and experimental
values from Ref.\cite{R}.} \\ 
\multicolumn{15}{l}{\small $^\mathrm{e}$ Sample and experimental
values from Ref.\cite{G1,G2}. } \\ 
\multicolumn{15}{l}{\small $^\mathrm{f}$ $E_{con}$ is obtained
experimentally from the shift between QW and QWR exciton lines.}  \\
\multicolumn{15}{l}{\small The $E_b$ is obtained indirectly from
experimental measurement of the QWR exciton line and one-particle
calculations. } \\ 
\multicolumn{15}{l}{\small $^\mathrm{g}$ Results of calculations from
Ref.\cite{Nonvar1}. } \\
\multicolumn{15}{l}{\small $^\mathrm{h}$ Results of calculations from
Ref.\cite{Nonvar2}.} \\
\multicolumn{15}{l}{\small $^\mathrm{i}$ Results of variational
calculations from Ref.\cite{Var1}.} \\
\multicolumn{15}{l}{\small $^\mathrm{j}$ Results of variational
calculations from Ref.\cite{Var2}.} \\
\multicolumn{15}{l}{\small $^\mathrm{k}$ Results of variational
calculations from Ref.\cite{Var3}. } 
\end{tabular}
\end{center}
\label{table}
\end{table}
\end{landscape}

Our results for the confinement energy of the ground state exciton
$E_{con}$ agree very well with experimental values for samples S1, N2
and N4 to an accuracy of 1\%, 6\% and 8\% respectively. This is indeed
very good agreement taking into account the strong inhomogeneous
broadening of the peaks they present. The spectral linewidth of the
photoluminescence peaks according to the authors is around 15 meV
which corresponds to a thickness fluctuation of about 3 {\AA} for N2
and N4
\cite{N2N4}. For the S1 and S2 samples the authors estimate the
experimental error due to the inhomogeneous broadening as 2 meV.
Agreement between our calculations and experiment is not as good for
the S2 sample but for this case additional effects are present.
For example, AlAs barriers give much less smooth interfaces than the
lower Al fraction samples and this is not taken into account in our
model. There is also very good agreement (better then 7\%) between our
results and the experimental measurement \cite{R} for asymmetric wire
R. The earlier $E_{con}$ published by this group for the symmetric
structure W is probably slightly overestimated.

All calculations published to date use the effective mass
approximation model for the heavy-hole exciton. Values of potential
barriers used in the calculations vary depending on the
publication. We have examined the influence of these differences on
the final results (see Section \ref{accuracy}). Both binding and
confinement energies can differ by approximately 0.5 meV

There are only two calculations published for the confinement energy.
They are based on variational methods and were performed only for
sample W. Variational method 2 \cite{Var2} uses a wave function which
takes into account correlation in all spatial direction and the
agreement with our results is very good for the confinement energy but
not so good for the binding energy.

The variational method proposed by Kiselev et al. \cite{Var3} and
denoted here by ``3'' has a trial wave function which has only $z$
dependence in the correlation factor. Their binding energy for the
sample W differs by only 1 meV from our result but their value for the
confinement energy differs from ours. They
perform calculations of the binding energy for the whole range of
well widths, $Dx$, from 10-70 {\AA}. This can be compared with our results
in Figure \ref{symEcEb}. Their calculations, like ours, give the
maximum for $E_b$ and $E_{con}$ for a well width of around
20 {\AA}. Their binding energy is a bit bigger than the one
from our calculations. They obtained a maximum of $E_{b}=18.6$ meV which
is 1.5 meV higher than our result. However, their confinement energy
$Econ_{max}=33.0$ meV differs by 7 meV from our result. Their values of
$E_{con}$ are probably overestimated. They use the variational
technique to calculate the quantum wire exciton energy but the quantum
well exciton energy is taken from some other calculations of excitons
in quantum wells performed using a different method and with different
parameters, thus errors may accumulate.

The variational method 1 \cite{Var1}, which uses yet another form of
trial wave function, has been applied to samples S1 and S2 to calculate the
binding energy $E_b$. It agree quite well with our and other accurate
methods.

The binding energy we obtain shows excellent agreement with another
non-variational calculations by Glutsch et al.\cite{Nonvar1} (see
Table \ref{table}). They calculated the binding energy only for
samples W, S1 and S2 and thus unfortunately the confinement energy
cannot be compared. The method presented in reference \cite{Nonvar2}
gives much lower values for the binding energy than all other methods.

Despite some small differences, all of the theoretical methods give
much smaller values for $E_b$ than the experimental estimates. One has
to bear in mind, however, that the ``experimental'' values for $E_b$
(quoted in the Table \ref{table}) are in fact derived from a
combination of experimental data and associated theoretical modelling,
with inherent uncertainties.  Our results come from direct
diagonalisation and are very well converged. Therefore we believe that
the experimental binding energies are, in some cases, considerably
overestimated. The real binding energy is thus smaller than has been
claimed and the biggest value for any of the structures manufactured
so far is 16.5 meV for samples S2 and N4.

\subsection{Optimisation of Confinement Energy for Experimental Realisation}
\label{opt}
In Section \ref{trends} we have shown that for symmetric
GaAs/Al$_{x}$Ga$_{1-x}$As wires the upper limit for the binding energy
is around 25 meV. We have also shown that in asymmetric structures,
the confinement energy is enhanced with respect to the symmetric forms
with comparable parameters but the binding energy of the exciton is
then lower than in the symmetric structures. The upper bound of 25 meV
is too low for the room temperature applications. There are some
indications (Section \ref{trends}) that
In$_{y}$Ga$_{1-y}$As/Al$_{x}$Ga$_{1-x}$As might be a good candidate.

Good quality GaAs/Al$_{x}$Ga$_{1-x}$As wires are, however, much easier
to manufacture that the In$_{y}$Ga$_{1-y}$As/Al$_{x}$Ga$_{1-x}$As
ones, which brings a considerable interest in optimising the
confinement energy for the GaAs/Al$_{x}$Ga$_{1-x}$As wires. The form
of the density of states for electrons and holes in 1D (see Chapter
\ref{Intr}) is very useful for low-threshold laser applications. 
Therefore structures in which excitons dissociate at room temperature
but remain in the wire would also be relevant for practical applications.

We collaborate with experimentalists from Bell-Laboratories to design
such lasers. The optimisation of the confinement energy of excitons is
a crucial issue in this design. We have shown in Section \ref{trends}
that in the asymmetric T-shaped wires, the confinement energy is
enhanced with respect to the symmetric forms, with comparable
parameters. We will try now to optimise the GaAs/Al$_{x}$Ga$_{1-x}$As
asymmetric wires within the well thickness and Al concentration range
accessible by experiment.

We perform calculations for structures shown in Figure \ref{asympot}
with a 140 {\AA}, 120 {\AA} and 100 {\AA} Stem well filled with 7
\%, 10 \% and 13 \% Al respectively. For all four different Stem wells
we vary the width of the Arm well for four different potential
barriers corresponding to 100 \%, 70 \%, 50 \% and 35 \% of Al
content.  The Arm well width is varied from 90 {\AA} to around the
value for which $E_{con}$ has a maximum. For each structure this
maximum corresponds to the case in which exciton energies in the Arm
and Stem quantum wells are equal. When the width of the Arm well is
decreased further the Arm well exciton energy increases, causing the
exciton to spread more into the Stem well and reducing the quantum
confinement.

\begin{figure}[htbp]
      \begin{center}  
        \leavevmode 
      \epsfxsize=16.0cm 
      \epsfbox{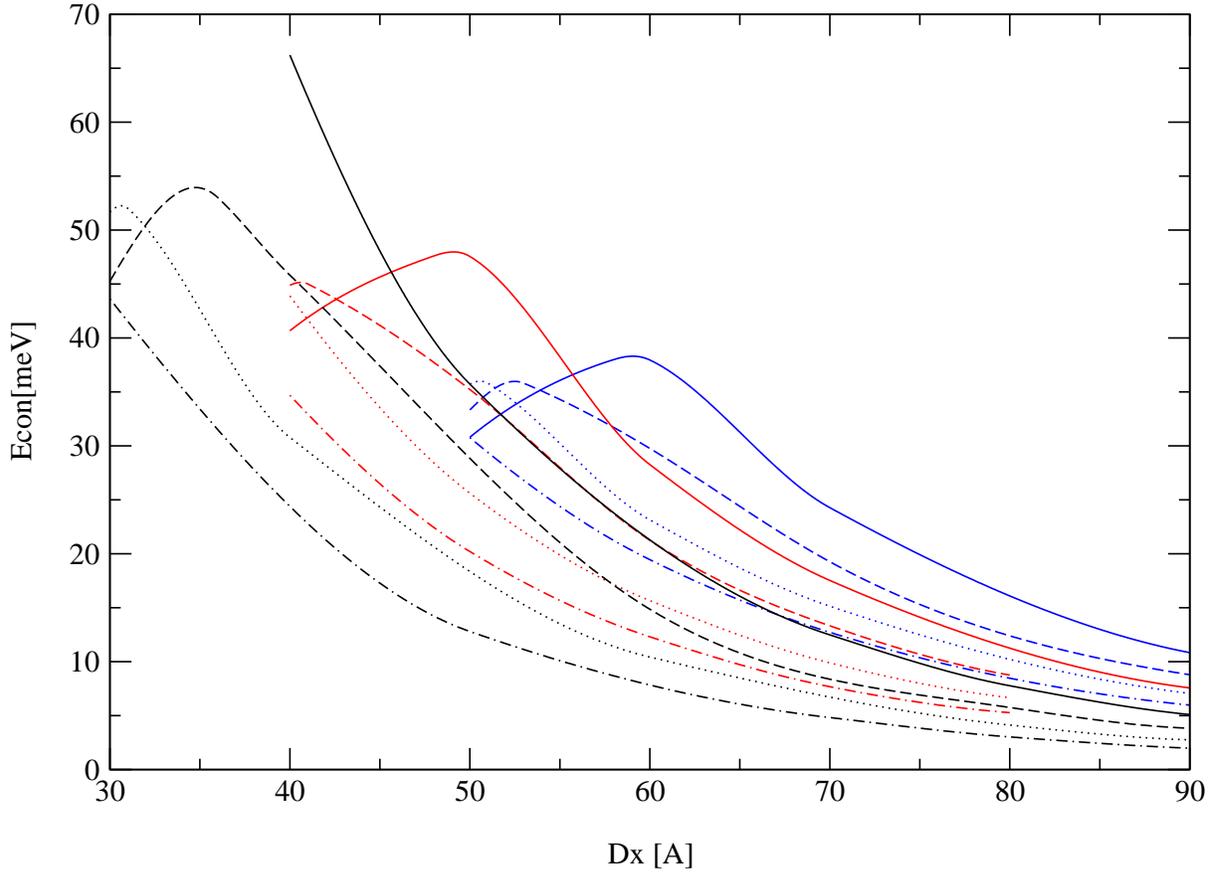} 
        \end{center}
 \caption{Confinement energy $E_{con}=E_{2Dexc}-E_{1Dexc}$ for asymmetric
wires as a function of the Arm well width $Dx$, for different Stem
wells and potential barriers. Blue curves correspond to the
140{\AA}/ 7 \% Stem well, red curves to the 120 {\AA}/ 10 \% and 
black curves to the 100 {\AA}/ 13 \% Stem well. Solid lines are for
100 \% Al barriers, dashed lines for 70 \% Al, dotted lines for 50
\% Al and dashed-dotted lines are for 35 \% Al barriers.}
\label{ak}
\end{figure}

The binding energy of excitons does not vary much within the studied
range and it changes from 17 meV for a wire which consists of a 100
{\AA}/ 13 \% Stem well, 40 {\AA} Arm well and 100 \% Al barriers to 12
meV for a wire with a 140{\AA}/ 7 \% Stem well, 90 {\AA} Arm well and
35 \% Al barriers. The values of $E_b$ for these wires are also
slightly smaller than for the corresponding symmetric structures (see
Figure \ref{nasEbEc} - upper panel). The confinement energy, however,
is much enhanced with respect to the symmetric forms (see Figure
\ref{nasEbEc} - middle panel) and strongly depends on its
parameters. The results of our calculations for $E_{con}$ are shown in
Figure \ref{ak}.
We now discuss the general trends for the confinement energy.

The highest confinement energies can be achieved for the 100 {\AA}/ 13
\% Stem well (black curves in Figure \ref{ak}), then for the 120
{\AA}/ 10 \% (red curves) and finally for the 140{\AA}/ 7 \% Stem
wells (blue curves). However the maximum value of $E_{con}$
corresponds to a different Arm well width for each different Stem
well. Notice that this maximum corresponds to a 30-40, 40-50, 50-60
{\AA} Arm well width for the 100 {\AA}/13 \%, 120 {\AA}/ 10 \%, and
140{\AA}/ 7 \% Stem well respectively. Thus although with a 100
{\AA}/13 \% Stem well we can achieve the highest confinement energies
we need to manufacture the thinnest Arm wells to achieve this. On the
contrary for wider Arm well the highest $E_{con}$ can be achieved with
the 140{\AA}/ 7 \% Stem wells.

It could be possible to introduce yet different types of Stem well to
increase $E_{con}$ even further. Unfortunately, the higher $E_{con}$
could only be achieved by using Arm wells thinner than 30 {\AA} which
would not be practical. The dependence of $E_{con}$ on the Al content
in potential barriers is the simplest one. $E_{con}$ always increases
with an increase of Al content.  

Although generally the value of the confinement energy is highest
for high Al contents and thin Arm wells both these features bring
additional experimental difficulties \cite{akiyama}:
\begin{itemize}
\item For higher Al concentration, yields of cleavage in CEO become
worse and interfaces get rougher which degrades the transport
properties.
\item Above 50\% AlGaAs becomes an indirect-gap semiconductor.
\item It becomes harder to make an optical wave guide for a laser
structure with higher Al concentration.
\item Thinner Arm-QWs cause higher thresholds in lasing.
\item In thinner QWs the wavefunctions tends to penetrate deeply into
AlGaAs, which may reduce the photoluminescence.  
\end{itemize}

In optimising wires for practical applications one needs to take into
account both the above mentioned difficulties and the gain in the
confinement energy. The detailed calculations of $E_{con}$ for a wide
range of structures are therefore very important. Our calculations are
being used to design quantum wire lasers which would operate at room
temperature. This work, in collaboration with Bell-Laboratories, is
still in progress.

\subsection{Accuracy of the Results}
\label{accuracy}
In our method the one-particle energies and wave functions are
calculated first. The one-particle energies are very well converged
with respect to all the variables such as unit cell size, number of
points on the grid and the number of plane waves to an accuracy of 0.1
meV. We use on average as many as 160 000 plane waves which
corresponds to $400 \times 400$ points on the grid ($200 \times 200$
in the small unit cell). We obtain excellent agreement between our
energies for the single electron and hole and those obtained by
Glutsch et al.\cite{Nonvar1}. For the 70 {\AA}, $x$=0.35 symmetric
quantum wire we obtain $E_e=47.09$ meV and $E_h=7.47$ meV while their
results are $E_e=47.2$ meV and $E_h=7.5$ meV. According to our
calculations there is only one electron state confined in the wire and
its confinement energy $E_{2D-1D}$ (i.e., the difference between
well-like and wire-like electron states) is 9 meV. This is in very
good agreement with other methods. L. Pfeiffer et al. \cite{calc} using
eight band $\vec{k} \cdot \vec{p}$ calculations obtained a confinement
energy of 8.5 meV for the same structure. Kiselev et al. \cite{Var3}
using the so-called free-relaxation method obtained approximately the
same value of 9 meV.

These one-particle wave functions are then used as a basis set for the
two-particle calculations. The $E_{1Dexc}$ is very well converged with
respect to the number of points on the grid (as for the one-particle
calculations), and with the size of the basis set (see Appendix
\ref{app1}, Tables \ref{aptable1} and \ref{aptable2}). Convergence is
usually achieved with about $20 \times 20 \times 20$ (8000) basis
functions. In order to minimise finite size effects we use quite big
unit cells (from 43 times the well width, $Dx$ for very thin wires (10
{\AA}) to 7 times $Dx$ for the 80 {\AA} wire). The exciton energy
$E_{1Dexc}$ is converged to within about 0.2 meV and $E_{2Dexc}$ to
within 0.3 meV which gives an accuracy for $E_b$ of about 0.3 meV and
for $E_{con}$ of about 0.5 meV.

The other problem which can influence the accuracy of the results is
the uncertainty associated with the input parameters. The electron and
hole masses as well as the dielectric constant are standard but the
potential barriers vary a lot depending on the publication. We have
found quite different values of the potential offsets for the same
material interfaces in the literature. We have examined the influence
of this uncertainty on the final results by performing calculations
for the extrema of the sets of parameters found. Both binding and
confinement energies can differ by approximately 0.5 meV.

For the parameters that we are using, the results are converged to
within 0.3 meV for the binding and 0.5 meV for the confinement
energies. However, one needs to remember that these parameters are not
well calibrated and this could lead to an additional error in both
energies of about 0.5 meV.

The first few (4 in the case of Fig \ref{asymspec}) excited states of
the wire which are below the 1Dcon and the 2Dexc are discrete,
quasi-1D excitonic states and are converged to within 1 meV.
Convergence of the higher states in the continuum is more
complicated. Because our system is finite we obtain only a sampling of
the continuum states. When we increase the unit cell size we
automatically calculate more states within the same energy region and
they do not have a one to one correspondence with the states
calculated using a smaller unit cell. The new states appear in between
the old ones, with smaller oscillator strengths so that the total
oscillator strength is conserved. When the Gaussian broadening of a 4
meV (FWHM) is added to the spectra then for a sufficiently big unit
cell the broadened spectrum is independent of the unit cell, thus
convergence is reached (see Appendix \ref{app1}, Figures
\ref{fig:ap_dens} and \ref{fig:ap_os}). The spectra shown in Figs
\ref{symspec} and \ref{asymspec} are converged in the sense that the
continuum is accurately sampled on the scale of 4meV.

\section{Conclusions}

We have performed an exact diagonalisation within a finite basis set
of the Hamiltonian which describes an interacting electron-hole pair
in a T-shaped quantum wire. We have obtained the ground and excited
state energies and wave functions for this system. The first group of
excited states shows an $s$-like excitonic character where the
electron is localised in the wire but is bound to the hole which
spreads into one of the wells. Due to the fact that the electron and
hole are not localised in the same region, we have a group of low
oscillator-strength states just above the ground state. This group is
followed by a number of states with large oscillator strength which
are 2D excitonic states scattered on the T-shaped intersection. The
excitonic lasing from one of those states has been experimentally
observed \cite{R}. We have also performed a detailed study of the
exciton binding and confinement energies as a function of the well
width and Al molar fraction for symmetric and asymmetric wires. The
highest binding energy in any structure so far constructed is
calculated to be 16.5 meV which is much smaller than previously
thought. Our results have shown that for optimised asymmetric wires,
the confinement energy is enhanced but the binding energy is slightly
lower with respect to those in symmetric wires. For
GaAs/Al$_{x}$Ga$_{1-x}$As wires we have obtained an upper limit for
the binding energy of around 25 meV in a 10 {\AA} wide GaAs/AlAs
structure which suggests that other materials need to be explored in
order to achieve room temperature applications.
In$_{y}$Ga$_{1-y}$As/Al$_{x}$Ga$_{1-x}$As might be a good candidate.

\chapter{ Two-mode Excitonic Lasing in T-shaped Quantum Wires } 
\chset{Two-mode excitonic Lasing in T-shaped quantum wires}
\label{bell}

{\it In this Chapter we present the experimental observation of
two-mode lasing in asymmetric, T-shaped quantum wires \cite{R}. Under
strong excitation the simultaneous lasing from two levels in the
structure is achieved. We then apply the numerical calculations
described in Chapter \ref{PRB} for structures used in this experiment
and identify the origin of the two laser modes.}

\section{Experimental Set-up}

The laser structure studied experimentally consists of two
intersecting quantum wells fabricated by cleaved edge overgrowth
molecular beam epitaxy (MBE). One set of quantum wells called Stem
wells are grown on a (001) GaAs substrate, and then in a second MBE
growth an intersecting Arm quantum well is grown on the (110) crystal
face exposed after an in situ cleave of the GaAs wafer in the growth
chamber (see Fig. \ref{asympot}). Previously reported work studied a
symmetric configuration in which Arm and Stem quantum wells had the
same width \cite{W}. The work described here is in an asymmetric
structure, where the Stem well is significantly wider but filled with
low percentage AlGaAs to compensate for the reduction in confinement
energy. The structure is composed of 20 Stem QWs, 140 {\AA} wide with
a 7 \% Al content intersecting a single 56 {\AA} or 60 {\AA} GaAs Arm
QW to form the T-QWRs. These 20-wire structures are embedded in a
multilayer arrangement that forms an optical cavity whose fundamental
mode concentrates the emitted light around the quantum wires. A
detailed description of the optical cavity can be found \cite{W}.
Along the wire axis, the 500 mm long laser cavity is limited by two
additional (011) cleaved surfaces that are left uncoated. The carriers
are resonantly excited by optically pumping the Stem quantum well
adjacent to the wire intersection. These carriers populate the Stem
wells and after phonon emission \cite{14}, they relax to wire states
and eventually recombine. The enlarged absorption length associated
with the wider stem wells in this sample results in much larger
carrier generation than in previous T-wire laser devices. This
manifests itself also in a x4 reduction of laser threshold power
compared with symmetric structures. In what follows, the excitation
power refers to that of the pump laser focused homogeneously over the
whole optical wire cavity area, and a maximum pump power of 500 mW
corresponds to a density of $\sim 2kW/cm^2$.

\section{Experimental Data}

\begin{figure}[p]
      \begin{center}
        \leavevmode
      \epsfxsize=8.5cm
      \epsfbox{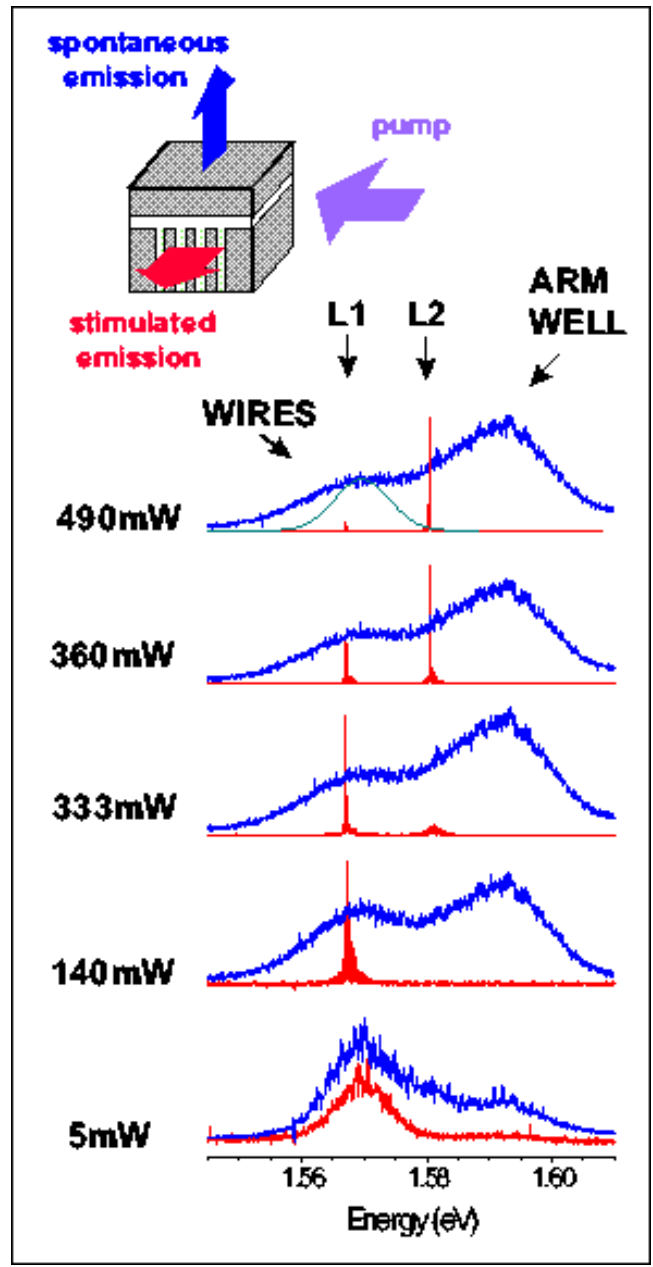}
        \end{center}
\caption{
Spontaneous (blue) and stimulated (red) emission of the
sample at T = 65 K for different excitation powers reflecting the
influence of the pump power on the transition between L1 and L2 lasing
lines. The spontaneous emission broadens to the high-energy side and
develops a long low-energy tail with increasing power. To expose these
effects a best fit of the QWR luminescence at the lowest power has
been added to the spectra at 490 mW. The band at 1.593 eV is due to
recombination associated with the 60 {\AA} arm QW.}
\label{fig:bell1}
\end{figure}

The effect of changes in photoexcitation intensity on lasing and also
on the spontaneous emission in these T-wire lasers is displayed in
Fig. \ref{fig:bell1}. 
Below threshold both the on- and off-axis spectra show the
luminescence from the excitonic ground state of the wires at 1.570 eV.
The luminescence from the Arm well appears as a weak shoulder in the
off-axis emission but becomes more prominent as a second peak at
1.593 meV as the photoexcitation energy is raised. At 140 mW the sample
is lasing in multimode with the gain centred at 1.567 meV (L1), an
energy which is very close to the excitonic ground state of the
wire. This photon energy is preserved for both kinds of emissions over
the whole range of excitation powers. At pump powers above 300 mW a
second lasing line (L2) appears and eventually dominates the laser
emission. Simultaneous lasing at two lines in a single cavity implies
a population inversion of more than one quantum state.

The positions of both laser lines remain largely constant, and L1
stays centred at an energy that is very close to that of the exciton
at low excitation power. Given the accuracy in the measurement of the
lasing line we can state that any shift is less than the spacing
between adjacent cavity modes (0.3 meV). The observations seem to imply
that laser gain displays excitonic behaviour even at the high
excitation powers.

\begin{figure}[p]
      \begin{center}  
        \leavevmode 
      \epsfxsize=10.0cm 
      \epsfbox{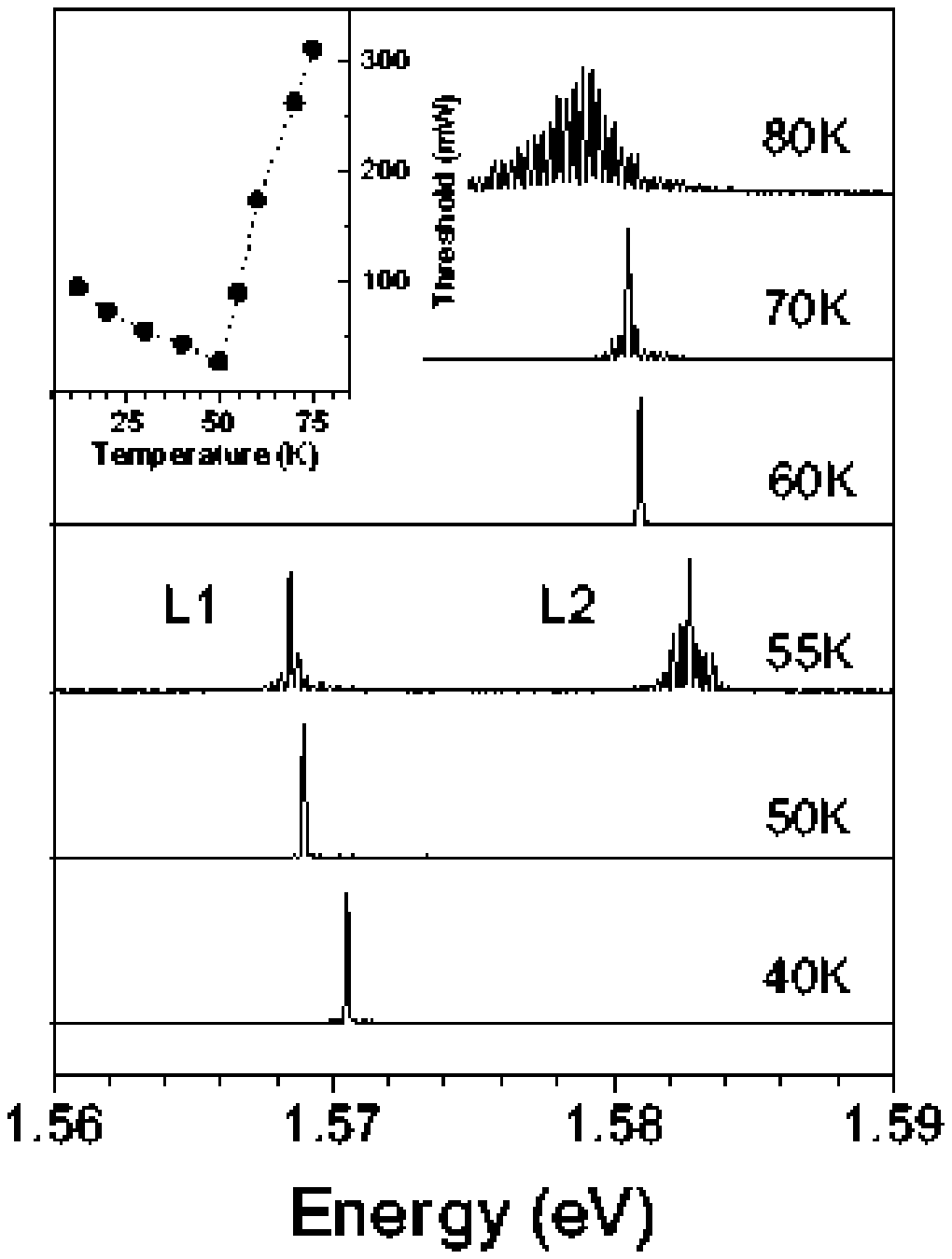} 
        \end{center}
 \caption{ Stimulated emission from the multiple QWR laser structure
        with resonant optical excitation of 500 mW resonant with
        exciton peak of the Stem well absorption. The system becomes
        bistable at 55 K and switches to a new line displaced 14 meV in
        energy for higher temperatures. Inset: evolution of the
        threshold power with temperature.}
\label{fig:bellnew1}
\end{figure}

The switching between the L1 and L2 laser lines shown in Fig. 
\ref{fig:bell1} has the striking temperature dependence displayed
in Fig. \ref{fig:bellnew1} \cite{R2}.
At 40 K the sample is lasing in a well defined single mode at 1.57
eV. In the temperature range 50 - 60 K, when the pump power exceeds
about 20 times threshold (see inset in Fig. \ref{fig:bellnew1}), the
drastic change in the laser emission occurs. The system becomes
bistable with emergence of the second line (L2), blue shifted 14 meV
from the fundamental line (L1). Raising the temperature further causes
the intensity of L1 to quickly diminish and the single mode lasing
from the L2 lines is obtained. After the L2 line is established the
threshold power begins to rapidly increase with temperature, causing
the emission to decay into a multimode operation that persists to 80 K.

It has to be stressed that with increasing temperature the L1 line
disappears, but with increasing pump power it saturates (when L2 turns
on) and then slowly diminishes in absolute intensity.

\begin{figure}[p]
      \begin{center}  
        \leavevmode 
      \epsfxsize=9.0cm 
      \epsfbox{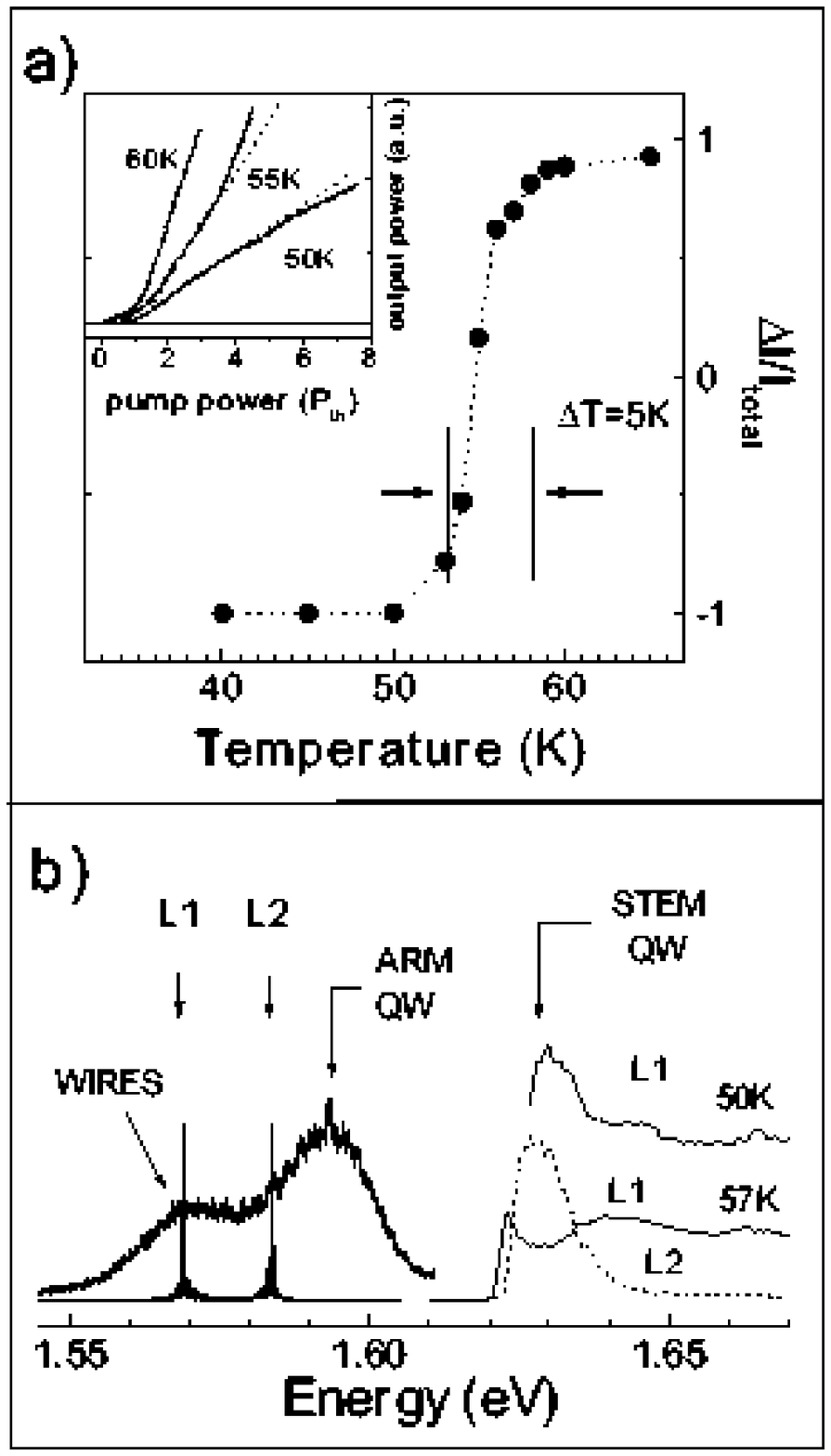} 
        \end{center}
 \caption{ a) Normalised difference of integrated intensities
        $(I_{L2}-I_{L1})/I_{Total}$ as a function of temperature in
        the transition region. Inset: Laser intensity versus pump
        power for three temperatures near the transitions (the pump
        power is normalised to the threshold value at each
        temperature). The straight dotted lines are guides to the
        eye. b) Spontaneous and laser emission at 57 K and the separate
        excitation spectrum of the two laser lines for two different
        temperatures. The spectrum at 57 K reveals the strong coupling
        between the two optical transitions associated with L1 and
        L2. }
\label{fig:bellnew2}
\end{figure}

Figure \ref{fig:bellnew2} presents the normalised difference of the
two integrated intensities and shows that switching abruptly takes
place within an interval $\Delta T \approx 5K$ around 55 K. At
temperatures above 55 K the intensity of L2 increases monotonically
with pump power, while L1 starts to diminish once both intensities
become comparable. With the onset of L2 an increase up to 33 \% in
the slope can be measured as a kink in the response curve, in
comparison with the gain saturation observed at lower temperatures
where L2 is absent (see inset in Fig \ref{fig:bellnew2}). These
characteristics are suggestive of a phase transition, where carriers
and photons are involved. A system of two lasing modes, represented
by two coupled oscillators, has a rich phase diagram which will be
expected to influence its photon statistics properties
\cite{two-mode}.

In the Figure \ref{fig:bellnew2} b the excitation spectra of both
lasing modes are depicted for two temperatures. At temperatures below
the transition (50 K), when only L1 is active, the excitation spectrum
reproduces the expected absorption of the Stem well. But when both
lasing lines are present, their intensities are strongly coupled as
evidenced by the excitation spectrum at 57 K. Once L2 is activated, it
competes with L1 to the point that the excitation spectrum of L1
develops a dip where L2 is maximum.

\begin{figure}[p]
      \begin{center}
        \leavevmode
      \epsfxsize=15.0cm
      \epsfbox{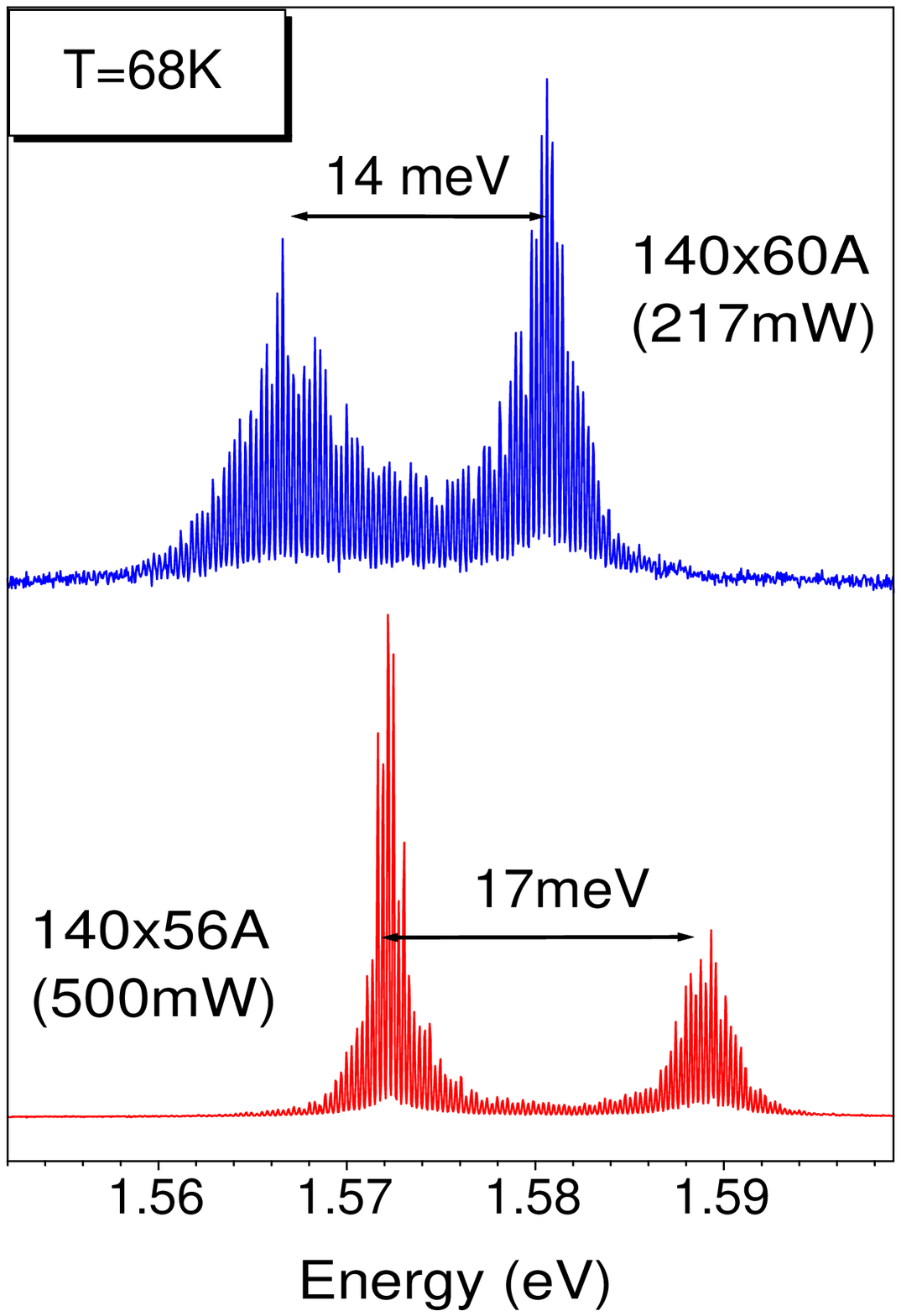}
        \end{center}
\caption{ Stimulated emission from the multiple QWR laser structure with
resonant optical excitation of 500 mW resonant with exciton peak of the
stem well absorption, for two samples with arm well widths of 60 {\AA}
and 56 {\AA}, otherwise identical. Both systems become bistable near
60 K and switch to a new line displaced 14 meV (17 meV) in energy for
higher temperatures.}
\label{bell2}
\end{figure}

While the L1 line clearly corresponds to the excitonic ground state of
the wires the origin of the L2 mode is not obvious. Since the binding
energy of the exciton in this geometry is approximately 14 meV, one
possibility might be exciton dissociation into itinerant quantum wire
states. To distinguish possible scenarios another T-wire laser was
studied with narrower Arm QW of 56 {\AA}, which should have a nearly
identical exciton binding energy, but a different energy level
structure. In this sample a substantially higher L1 - L2 splitting of 17
meV was observed (see Fig. \ref{bell2}).
The 20 \% increased shift is not consistent with that exciton breakup
hypothesis. 

Since the L2 transition is only observed at high excitation levels the
first suggestion was that L2 has entirely a many-body origin,
connected with the presence of an electron-hole plasma in the system.
Such persistence of excitonic effects in 1D to high carrier density
would be very remarkable.

Another possibility would be that the second lasing mode (L2)
corresponds to a higher excitonic state in the structure.

Sorting out these possibilities represents a renewed challenge as
there is no independent determination of the density.  Both scenario
suggest that an upper limit for the density is approximately 1 per
Bohr radius and thus is roughly of the order that one would expect for
the transition from e-h plasma to bound excitons.


We have performed calculations which suggest that the L2 line corresponds
to a higher excitonic state in the structure. We have first developed a
rate equation model for the two-mode laser, presented in Chapter
\ref{rate}, assuming that the laser action takes place for two
exciton-like states. This model can qualitatively reproduce the
experimental data provided that the oscillator strength times the density
of states for the upper excitonic state (L2 line) is higher than that for
the ground state (L1 line). Motivated by this result we have performed
detailed calculations for both structures studied in the experiment (56
{\AA} and 60 {\AA} Arm QWs) to identify the excitonic states which
correspond to the two laser modes.

\section{Calculations}

\begin{figure}[p]
      \begin{center}
        \leavevmode
      \epsfxsize=13.0cm
      \epsfbox{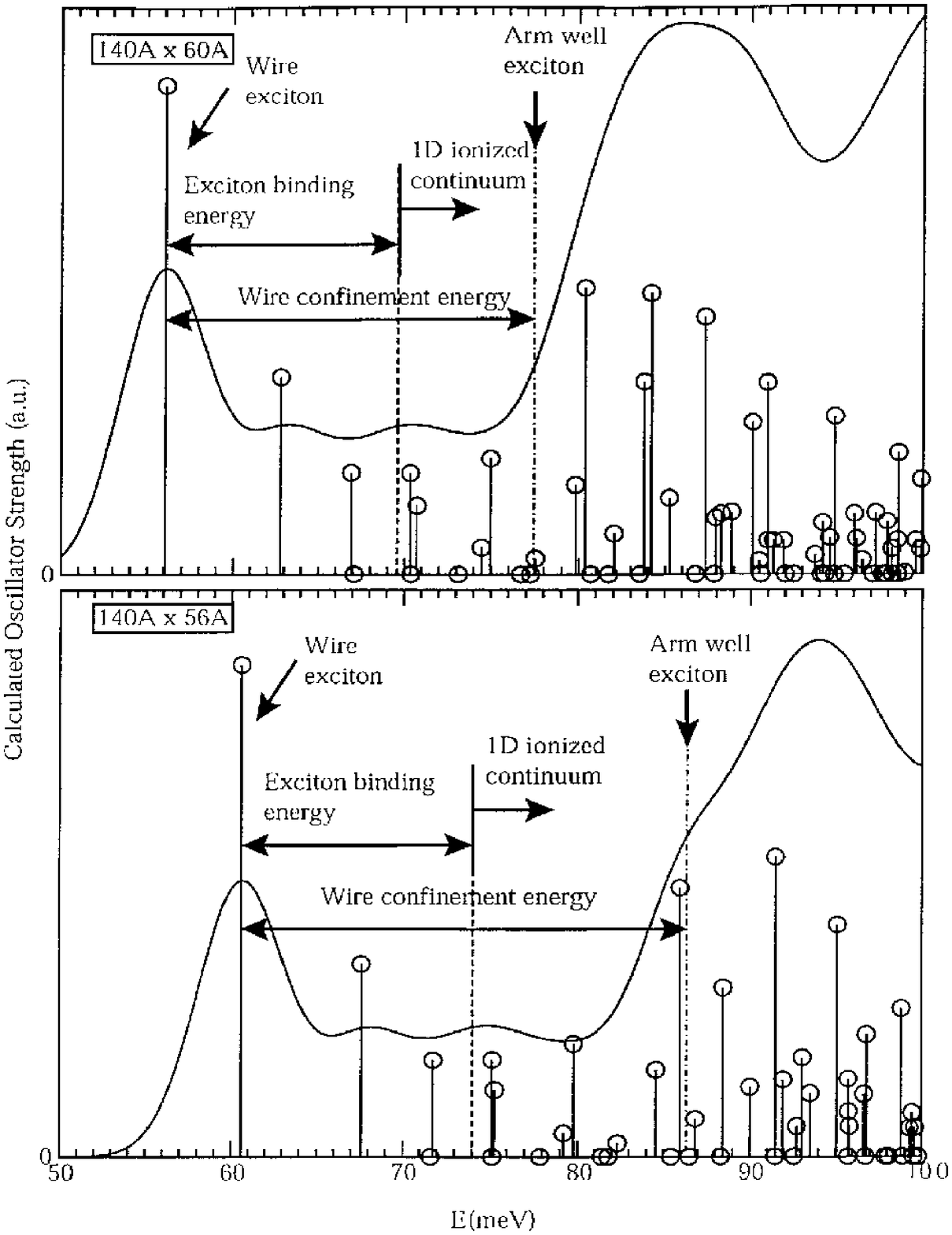}
        \end{center}
\caption{ Calculations of the energies and oscillator
strengths for ground and excited states of an electron-hole pair in
the same two geometries as experiment. The onset of the 1D continuum
of electron-hole plasma states, and the 2D exciton in the arm well are
indicated. The solid line is produced by a Gaussian broadening with 3
meV HWHM, comparable to the inhomogeneous linewidth in the
experimental system, and is meant as a guide to the eye. The peak in
the oscillator strength above the 2D exciton edge arises because
interactions with the wire perturb nonzero momentum excitons and make
them optically active.}
\label{bell3}
\end{figure}

We have performed numerical calculations of a single electron-hole
pair in the T-shaped geometries of the two experiments to explore how
the geometric changes might affect the spontaneous luminescence. To
this end, we performed (see Chapter \ref{PRB} and reference
\cite{marz}) exact diagonalisations to determine not only the ground
but also the excited states of the excitonic system, with results
shown in Fig \ref{bell3}.
The results for the two cases are similar: the ground state 1D exciton
is bound 14 meV below the one-dimensional particle-hole continuum, and
confined by 22 (26) meV in comparison to the 2D exciton in a 60 (56)
{\AA} arm well. There is an absolute shift of 3.5 meV in the
1D-exciton energy between the two structures, in agreement with
experiment. We also find several excited states of the bound 1D
exciton. States above the edge marking the free particle-hole
excitation of the wire are of course in a continuum, and the discrete
spectrum above 65 (69) meV is due to the finite box size of our
calculations. We note that there is a broad peak of oscillator
strength lying slightly above the 2D quantum well energy. Inspection
of the wavefunctions of the large oscillator-strength states reveals
that they are principally quasi-2D-exciton-like states strongly
perturbed by the quantum wire potential, resonant with (and thus
decaying into) the continuum states of the wire. The oscillator
strength of the 1D-continuum states themselves is weaker.

While these calculations are valid only for low density, the absence
of substantial shifts in the luminescence spectrum with pumping
suggests that the calculations can be used as a guide to the
assignment of the transitions. L2 lies near but below the energy
marking the onset of 2D excitons in the Arm well. Nevertheless, the
shift of L2 from 14 to 17 meV between the 60 and 56 {\AA} structures
mirrors the calculated change of the exciton confinement energy in the
wire from 22 to 26 meV, indicating that L2 may have well-like
character.

Since the L2 transition is only observed at high excitation levels, it
is possible that the 8-9 meV shift of the L2 line from the
calculated 2D exciton edge may be a redshift induced by screening by
the occupied 1D excitons. 

Very recent experiments \cite{loren} give an additional support for
the hypothesis that the L2 line is associated with the well excitons
in the vicinity of the wires. With a reduction of a number of wires in
the sample the threshold power for L1 lasing goes up while the
threshold for L2 remains constant.

\section{Conclusions}

We have presented the experimental observation of two-mode lasing in
asymmetric, T-shaped quantum wires \cite{R}. Under strong excitation
the simultaneous lasing from two levels in the structure was
achieved. We have applied the numerical calculations described in
Chapter \ref{PRB} for both structures, with 60 {\AA} and 56 {\AA} Arm
quantum well, used in this experiment to identify the origin of the two
laser modes.

We have obtained an absolute shift of 3.5 meV in the 1D-exciton energy
between the two structures, in agreement with experiment. We have also
found several excited bound 1D exciton states but there is no
evidence for emission from these states. Shift of L2 from 14 to 17 meV
between the 60 and 56 {\AA} structures mirrors the calculated change
of the exciton confinement energy in the wire from 22 to 26 meV,
suggesting that L2 has a well-like character, though the position of L2
is somewhat below the onset of 2D excitons in the arm well. Early
suggestions that L2 is associated with emission from 1D e-h plasma are
not substantiated.

\chapter{Rate Equation Model for a Two-Mode Laser}
\chset{Rate Equation Model for a Two-Mode Laser}
\label{rate}

{\it In this Chapter we develop a rate equation model for a two-mode
laser. Assuming that the laser action takes place for two different
exciton-like states, we can reproduce the same qualitative behaviour
of the laser, as observed in the experiment described in Chapter
\ref{bell}. The results of this Section support the hypothesis that
the second lasing mode (L2) is associated with the higher excitonic
state in the structure. The calculations of the spectra for
experimentally studied wires (see Chapter \ref{bell}) suggested that this
higher excitonic state has a well-like character.  }

\section{Experimental Motivations}

In Chapter \ref{bell} we have described in detail the experimental
observation of a two-mode lasing in asymmetric, T-shaped quantum
wires. The first lasing mode comes from the ground state exciton while
the nature of the second lasing line is not known. Determination of
the origin of this line represents a renewed challenge as there is no
independent determination of the density which might be very high, of
the order that one would expect for the transition from e-h plasma to
bound excitons. Thus many-body effects might be important.

The laser shows very interesting behaviour which might help to
determine the origin of the second lasing mode. There is a switching
between the two laser modes as the temperature or the pump power is
changed. At low pump powers and low temperatures only the first laser
line (L1) is present. There is a quite wide range of pump intensities
and a very narrow range of temperatures for which the laser works in
multimode in which both modes are lasing. At high powers or high
temperatures the second lasing line, L2, dominates. It has to be
stressed that with increasing temperature the L1 line disappears, but
with increasing pump power it saturates (when L2 turns on) and then
slowly diminishes in absolute intensity (see Figures \ref{fig:bell1}
and \ref{fig:bellnew1}). The crossover as the temperature is changed
in the experiment takes place within a very narrow interval of about 5K
(Figure \ref{fig:bellnew2}). With the onset of L2 an increase of up to 33
\% in the slope can be measured as a kink in the response curve, in
comparison with the gain saturation observed at lower temperatures
where L2 is absent (see inset in Fig \ref{fig:bellnew2})

We need to find a model for a two-mode laser, which could explain such
behaviour. In this chapter we assume that the second lasing mode is
also associated with an exciton-like state, higher in energy than the
ground state. 

\section{Model}

\begin{figure}[htbp]
\begin{center}
\includegraphics[width=14cm]{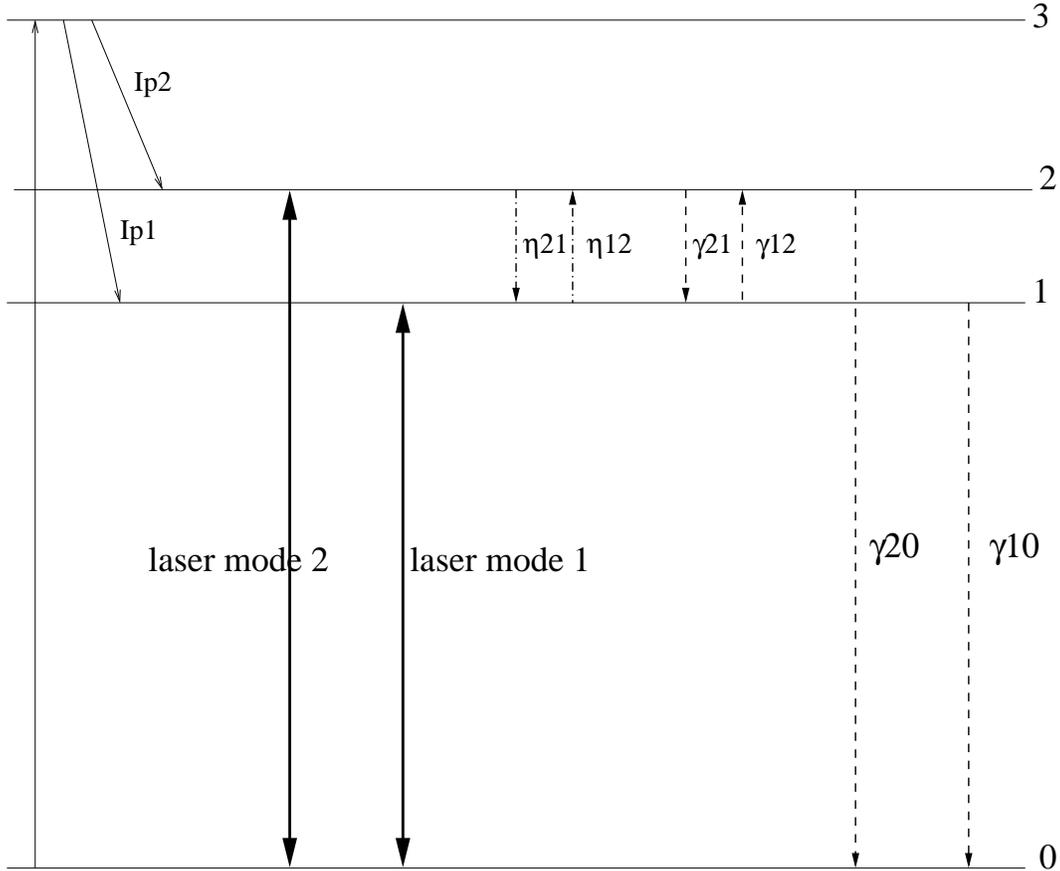}
\end{center}
\caption{Sketch of the model for a two-mode laser system. Various
	levels and the transitions between them are shown. Bold solid
	lines correspond to laser actions, dashed lines correspond to
	photon spontaneous emission and absorption while dotted-dashed
	lines correspond to phonon emission and absorption.}
\label{fig:diagram}
\end{figure}

The model we consider in this Chapter is schematically shown in Figure
\ref{fig:diagram}. Levels ``0'' and ``3'', shown in the diagram
\ref{fig:diagram}, correspond to the valence and the conduction band
of the semiconductor, respectively. Level ``1'' is a ground state
exciton while level ``2'' is a higher excitonic state in the
structure. Since excitons cannot be packed closer than around a Bohr
radius the maximum occupancy of levels ``1'' and ``2'' is limited and
given by excitonic density of states for both levels, respectively. We
call these maximum occupancies $N_1$ for level ``1'' and $N_2$ for level
``2''.

External excitations create free electrons and holes in the
semiconductor. This corresponds to pumping the electrons from the valence
(level ``0'') to the conduction (level ``3'') bands. Then they
populate the two excitonic states, levels ``1'' and ``2'', with rates
$I_p^{(1)}$ and $I_p^{(2)}$, respectively. The first laser action (L1
line) takes place between level ``1'' and level ``0'', while the second
laser mode (L2 line) is between levels ``2'' and ``0''. 

There are also other transitions present in the system. Phonons and
thermal photons would cause transitions between levels ``2'' and ``1''
since the energy difference between these levels is only 14 meV. We
introduce $\gamma_{21}$ and $\gamma_{12}$ as emission and
absorption rates, respectively, of thermal photons and $\eta_{21}$ and
$\eta_{12}$ emission and absorption rates of phonons. Absorption and
emission rates are connected through the Boltzmann factors
\begin{align}
\gamma_{12}&=\gamma_{21}e^{-\frac{\Delta}{kT}} &
\eta_{12}&=\eta_{21}e^{-\frac{\Delta}{kT}},
\label{eq:Boltzman}
\end{align}
where $\Delta$ is the energy difference between the two lasing modes.
We also consider spontaneous emissions of photons to modes different
from the two lasing modes for both lasing levels with rates
$\gamma_{20}$ and $\gamma_{10}$, respectively.

This model combines two standard rate equations models for
the semiconductor laser and for the atomic laser. The levels ``0'' and
``3'' are alike in the semiconductor model where the maximum occupancy
of the levels are not imposed while levels ``1'' and ``2'' are the
same as in atomic lasers characterised by the parameters $N_1$ and
$N_2$ which involve saturation of the levels.

Without detailed calculations we can see that raising the temperature
thermally depopulates level ``1'' into level ``2''.  A strong pumping
also increases the population of level ``2'' over level ``1'',
provided the trapping from level ``2'' into level ``1'' is relatively
inefficient in comparison to the rate of emission from level ``2'',
and provided also that the population of level ``1'' can saturate.

Introducing the abbreviation $r_{21}=\gamma_{21}+\eta_{21}$ and using
equation (\ref{eq:Boltzman}) we obtain, from a detailed balance, the
rate equation for the population of the higher excitonic state (level
``2''), $n_2$
\begin{multline}
\label{eq:n2}
\frac{dn_2}{dt} = Ip^{(2)}(N_2-n_2)-r_{21}n_2(N_1-n_1)+
                    r_{21}e^{-\frac{\Delta}{kT}}n_1(N_2-n_2) \\
  -\gamma_{20}n_2-n_2(1+S_2)g_2 +(N_2-n_2)S_2g_2.
\end{multline}
The first term in equation (\ref{eq:n2}) corresponds to the pumping of
level ``2'' and is proportional to the pump power and the
difference between the maximum number of excitons, $N_2$, and the current
number of excitons in this state, $n_2$. The second term, which
describes transitions from level ``2'' to level ``1'', caused
by phonons or thermal photons is proportional to the number of
excitons in level ``2'' and to the difference between the maximum
number of excitons in level ``1'', $N_1$, and the current number of
excitons in this state, $n_1$. Similarly the third term describes the
transitions from level ``1'' to level ``2 '', caused by phonon
or thermal photons. The fourth term corresponds to the spontaneous
emission of photons to modes other than the lasing modes. Finally, the
fifth term describes a stimulated emission of photons to the L2 lasing
line and is proportional to the number of excitons in level ``2'',
$n_2$, and to the factor 1+$S_2$, where $S_2$ is a number of photons in
the L2 mode. The sixth term gives the stimulated absorption of photons
from the L2 lasing mode.

Similarly, the rate equation for the number of excitons in the ground
state (level ``1''), $n_1$ is
\begin{multline}
\label{eq:n1}
\frac{dn_1}{dt}  =  Ip^{(1)}(N_1-n_1)+r_{21}n_2(N_1-n_1)-
                    r_{21}e^{-\frac{\Delta}{kT}}n_1(N_2-n_2) \\
   -\gamma_{10}n_1-n_1(1+S_1)g_1+(N_1-n_1)S_1g_1,
\end{multline}
where $S_1$ is the number of photons in the L1 lasing mode.

The rate equation for the photon number in the L1 mode is
\begin{equation}
\label{eq:rateS2}
\frac{dS_2}{dt}=-\kappa_2S_2+(n_2-(N_2-n_2))S_2g_2,
\end{equation}
while in the L2 mode is
\begin{equation}
\label{eq:rateS1}
\frac{dS_1}{dt}=-\kappa_1S_1+(n_1-(N_1-n_1))S_1g_1,
\end{equation}
where $\kappa_1$ and $\kappa_2$ are the cavity decay rates for the L1
and the L2 lasing mode, respectively.

\section{Steady State Behaviour}

We are now interested in a steady state behaviour of these four
coupled equations (\ref{eq:n2}) - (\ref{eq:rateS1}). We need to
consider four different regimes:

1. $S_1>0$ and $S_2>0$, both modes are lasing.

2. $S_1>0$ and $S_2=0$, 1st mode is lasing, 2nd is not lasing.

3. $S_1=0$ and $S_2>0$, 1st mode is not lasing, 2nd is lasing.

4. $S_1=0$ and $S_2=0$, both modes are not lasing.

Experimentally all four regimes were observed in the same structure
for different temperatures and pump powers (see Chapter \ref{bell}).
It can be shown that, to have all four regimes in the same structure in
our model, the laser parameters have to satisfy the following three
conditions
\begin{equation}
A = \frac{N_1}{2}-\frac{\kappa_1}{2g_1} > 0
\label{eq:cond1}
\end{equation}
\begin{equation}
C= \frac{N_2}{2}-\frac{\kappa_2}{2g_2} > 0
\end{equation}
\begin{equation}
\label{eq:cond3}
\frac{N_2g_2}{\kappa_2} > \frac{N_1g_1}{\kappa_1}.
\end{equation}
If we assume that the cavity decay rates for both laser modes are the
same $\kappa_1=\kappa_2$, the condition (\ref{eq:cond3}) means that the
oscillator strength times the density of states for level ``2''
has to be larger than that of level ``1''.  It is very unlikely that
the oscillator strength for any higher excitonic state would be larger
than that of the ground state, but it is very probable that the density
of states would be larger for the higher state and thus the condition
would be satisfied.

In the Chapter \ref{bell} we argued that the second lasing line is
connected with quasi-2D-exciton-like states strongly perturbed by the
quantum wire potential, resonant with (and thus decaying into) the
continuum states of the wire. The density of such states would be much
bigger than that of the ground state excitons (see Figure
\ref{bell3}).

Let us now consider in more detail the four different regimes 
as the temperature and pump power are changed.
\begin{itemize}
\item Multimode operation of the laser: $S_1>0$ and $S_2>0$. Using
equations (\ref{eq:n2}) - (\ref{eq:rateS1}) we can calculate the
intensities of both laser lines in a steady state, assuming that both
photon modes are occupied. Introducing abbreviations
\begin{equation}
B=\frac{N_1}{2}+\frac{\kappa_1}{2g_1},
\end{equation}
and
\begin{equation}
D= \frac{N_2}{2}+\frac{\kappa_2}{2g_2},
\end{equation}
the intensities of the modes will be 
\begin{equation}
S_2=\frac{r_{21}}{\kappa_2}(BCe^{-\frac{\Delta}{kT}}-AD+C\frac{Ip^{(2)}}
{\gamma_{21}}-\frac{\gamma_{20}+g_2}{\gamma_{21}}D),
\label{eq:S2}
\end{equation}
\begin{equation}
S_1=\frac{r_{21}}{\kappa_1}(-BCe^{-\frac{\Delta}{kT}}+AD+A\frac{Ip^{(1)}}
{\gamma_{21}}-\frac{\gamma_{10}+g_1}{\gamma_{21}}B).
\label{eq:S1}
\end{equation}
Notice that the first two terms in equation (\ref{eq:S2}), out of
which only the first one is temperature dependent in the whole
equation, have the same magnitude but opposite sign to the analogous
terms in equation (\ref{eq:S1}). These terms cause the competition
between the two modes as the temperature is changed. At low
temperatures the first two terms in equation (\ref{eq:S2}) would be
negative, and when the pumping (the third term) is too low to
compensate them and the losses described by the fourth term, the L2
line will not be present. Similarly, from equation (\ref{eq:S1}) we
can see that at high temperatures the L1 line will not be present. As
the temperature is changed, the rate equation model predicts the same
switching of modes, as was experimentally observed (see Chapter
\ref{bell}).

\item At low temperatures and pumping rates, when $S_1>0$
and $S_2=0$, the intensity of the L1 lasing line will be
\begin{equation}
S_1 =
\frac{AIp^{(1)}-B(\gamma_{10}+g_1)}{\kappa_1}+\frac{N_2}{\kappa_1}*
\frac{AIp^{(2)}-B(\gamma_{20}+g_2)e^{-\frac{\Delta}{kT}}}{A+
Be^{-\frac{\Delta}{kT}}+\frac{Ip^{(2)}}{\gamma_21}+\frac{\gamma_{20}+g_2}
{\gamma_{21}}}
\label{eq:S1e}
\end{equation}
\item At high temperatures or pump powers, when $S_1=0$ and $S_2>0$, the
intensity of the L2 line would be
\begin{equation}
\frac{1}{\kappa_2}*(CIp^{(2)}-(\gamma_{20}+g_2)D+
\frac{CIp^{(1)}e^{-\frac{\Delta}{kT}}}{Ce^{-\frac{\Delta}{kT}}+D+
\frac{Ip^{(1)}}{\gamma_{21}}+\frac{\gamma_{10}+g_1}{\gamma_{21}}}.
\label{eq:S2e}
\end{equation}
\end{itemize}

Equations (\ref{eq:S2}) - (\ref{eq:S2e}) have quite complicated forms,
therefore it will be easier to study the behaviour of the system by
plotting the results for particular parameters. We developed this
model before the detailed calculations of spectra, described in Chapter
\ref{bell}, were obtained and thus not all values of input parameters were
available. We noticed, however, that the qualitative behaviour of the
system is not that sensitive to the input parameters, provided that the
conditions (\ref{eq:cond1}) - (\ref{eq:cond3}) are satisfied. We choose
the input parameters so that the temperature transition region is the
same as in the experiment. For different, but yet realistic, parameters
the qualitative behaviour would remain the same, but the transition
temperature would change. 

The most important parameter which determines the thermal coupling of
the two modes is $\Delta$. We use exactly the same value of 14 meV for
$\Delta$ as in the experiment. The exact values of other parameters
are very difficult to determine. In this work we use approximate
values, but realistic ones for these types of structures.  We normalise
the parameters so that the gains $g_1$ = $g_2$ = 1, $N_2/N_1$ = 100,
$\kappa_1/g$ = $\kappa_2/g$ = 8.81, $r_{21}/\gamma_{20}$ =
$r_{21}/\gamma_{10}$ = 10. We choose $I_p^{(1)}$ to be 0, which is a
physical assumption, because the exciton has to get into the well
first before it can get into the wire.

\begin{figure}[htbp]
\begin{center}
\includegraphics[width=15cm]{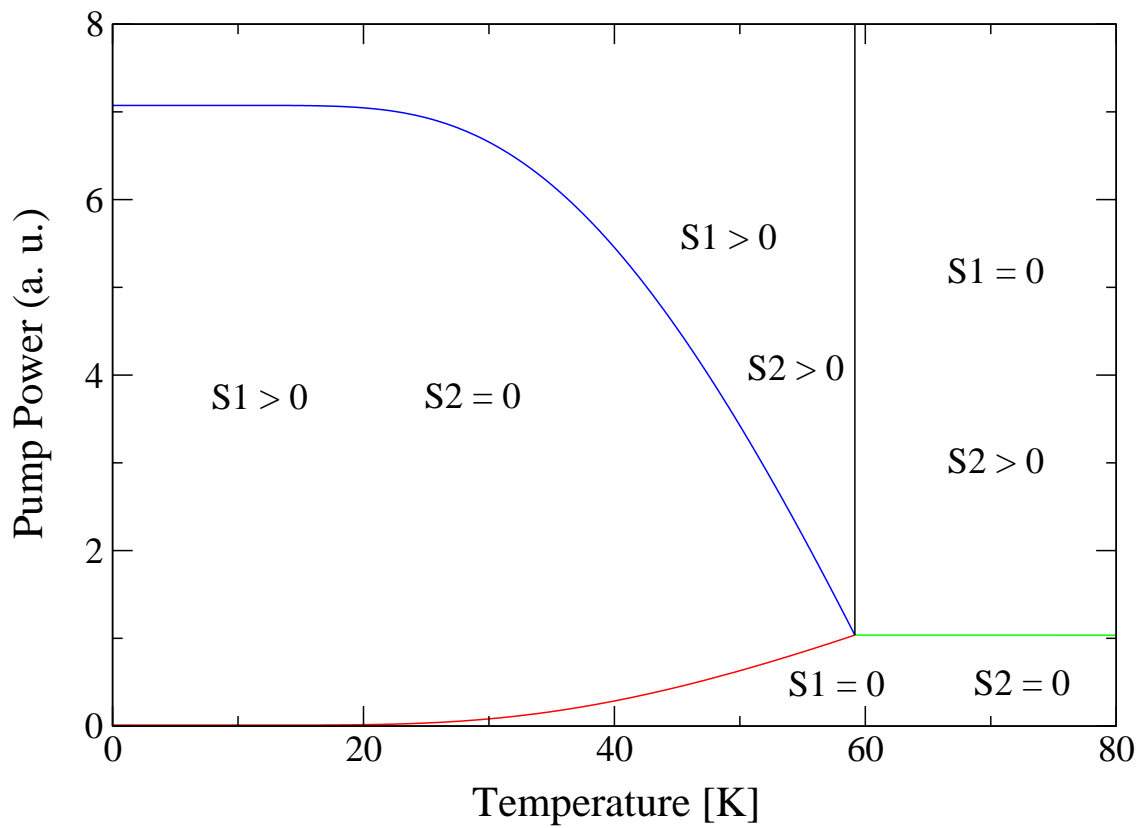}
\end{center}
\caption{Phase diagram for a two-mode laser. Phase boundaries are
shown as solid lines. The red (green) solid line corresponds to
transition between the lack of lasing and the appearance of the L1 (L2)
lasing mode. The blue line is a phase boundary between the presence of
the L1 mode only and the multimode operation (presence of both L1 and
L2 modes). The black line corresponds to the transition from
simultaneous lasing of L1 and L2 modes and the presence of the L2 line
only. }
\label{fig:rate_phase}
\end{figure}

In Figure \ref{fig:rate_phase} we present a phase diagram for our
system. The four different phases are marked in Fig
\ref{fig:rate_phase}. The phase diagram recovers the experimental
behaviour, shown in Figures \ref{fig:bell1} and \ref{fig:bellnew1}.
Let us first discuss the behaviour of the system at a fixed
temperature, lower than 59K. At very low pump powers there is no
lasing in the system. With the increase in the pump power the L1 laser
mode appears first. When the pump power is increased further, the L2
laser action starts and the system lases in multimode. The intensity of
the L1 line saturates when the L2 mode appears, while the intensity of
the L2 increases with the pump power (see Figure
\ref{fig:intensity2}). Thus the L1 line slowly diminishes in absolute
intensity in agreement with experiment (see Figure
\ref{fig:bell1}). At such high pump intensities, it is also possible
that the temperature of the T-wire would rise above the nominal
reservoir temperature, given on the graphs, which would additionally
suppress the intensity of the L1 line.

\begin{figure}[htbp]
\begin{center}
\includegraphics[width=12cm]{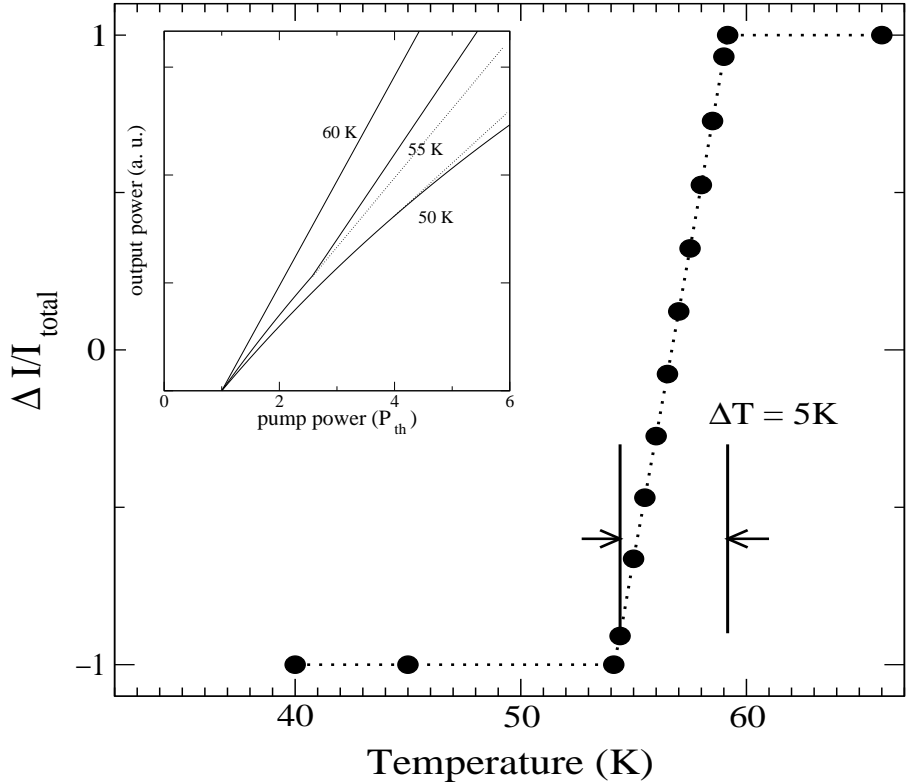}
\end{center}
\caption{ Calculated normalised difference of intensities
        $(I_{L2}-I_{L1})/I_{Total}$ as a function of temperature in
        the transition region. Inset: Laser intensity versus pump
        power for three temperatures near the transitions (the pump
        power is normalised to the threshold value at each
        temperature). The straight dotted lines are guides to the
        eye. This Figure should be compared to the experimental Figure
        \ref{fig:bellnew2}  }
\label{fig:rate_intens}
\end{figure}

\begin{figure}[h]
      \begin{center}
        \leavevmode
      \epsfxsize=9.0cm
      \epsfbox{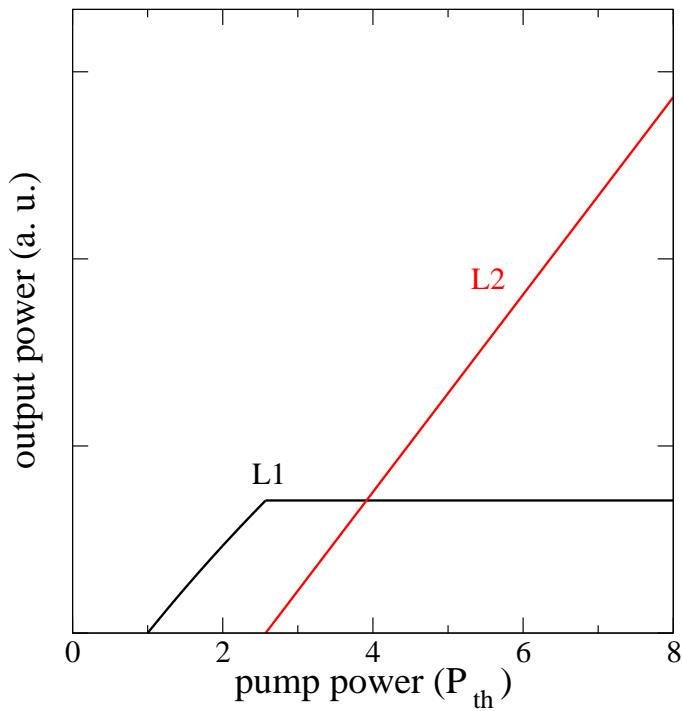}
        \end{center}
\caption{Intensity of the L1 (black curve) and the L2 (red curve)
lasing modes at 55 K. When the L2 line appears the total intensity has
a kink (see Figure \ref{fig:rate_intens}).}
\label{fig:intensity2}
\end{figure}

Similarly, when we fix the pump power at a value higher than the
threshold for lasing and change the temperature, we recover the
experimental behaviour, given in Figure \ref{fig:bellnew1}. At low
temperature the L1 mode is lasing, as the temperature is increased
both L1 and L2 modes are present, and finally the L1 line disappears
and only the L2 mode is present. This switching usually takes place
within a narrow interval of temperatures (see Figure
\ref{fig:rate_intens}).

In the inset of Figure \ref{fig:rate_intens} we present the laser
intensity versus pump power for three temperatures near the
transition.
With the onset of L2 an increase in the slope can be observed as a
kink in the intensity curve, in comparison with the gain saturation
observed at lower temperatures where L2 is absent in agreement with
experiment (see inset in Fig \ref{fig:bellnew2} a). The intensity
at 60 K comes from the L2 line while at 50 K from the L1 line only. At
55K both lasing lines contribute, as shown in Figure \ref{fig:intensity2}.

Figure \ref{fig:rate_intens} presents the normalised difference of the
two mode intensities and shows that, in agreement with experiment
(see Figure \ref{fig:bellnew2} a ), switching takes place within a
narrow interval. We choose parameters so that the switching takes
place within an interval $\Delta T \approx 5K$ around 55K. For
different, but still realistic, parameters the transition temperature
would be different but the qualitative behaviour would remain the
same. At temperatures above 55K the intensity of L2 increases
monotonically with pump power, while L1 starts to diminish once both
intensities become comparable.

\section{Conclusion}

We have shown, using a simple rate equation model, that the
experimental results described in Chapter \ref{bell} can be explained
assuming that the laser lines come from two different excitonic states
in the structure. The most important qualitative features of the model
that are necessary to explain the data are:
\begin{itemize}
\item the saturation of the states associated with the lasing lines,
\item the temperature-dependence of the thermal occupation of these
states, 
\item a relatively inefficient trapping from the higher
to the lower lasing state in comparison to the rate of photon emission
from the higher lasing state.
\end{itemize}

Using this model we have recovered: 
\begin{itemize}
\item the switching between the two modes
as temperature is changed, 
\item the saturation of the L1 line as the
pump power is increased, 
\item the region of a multimode operation,
\item experimentally observed behaviour of intensities of the two lasing
lines.
\end{itemize}

This very good qualitative agreement between the model and the
experiment suggests that we indeed have lasing from two different
excitonic states in the structure. Motivated by this agreement we
performed detailed calculations, presented in the Chapter \ref{bell},
to identify the L2 excitonic state.

\chapter{Summary and Future Directions}
\label{conT}
\section{Summary}
In this work we have developed a method to calculate electron-hole
states in quantum wires. We have included the single-particle
potential and the Coulomb interaction between the electron and hole on
an equal footing, and have performed exact diagonalisation of the
two-particle problem within a finite basis set. We have calculated
energies, oscillator strengths for radiative recombination, and
two-particle wave functions for the ground state exciton and around
100 excited states in a T-shaped wire. 

We have studied in detail the shape of the wave functions to gain insight
into the nature of the various states for selected symmetric and
asymmetric wires in which laser emission has been experimentally
observed. We have found
\begin{itemize}
\item That the first group of excited states shows an $s$-like
excitonic character where the electron is localised in the wire but is
bound to the hole which spreads into one of the wells. Due to the fact
that the electron and hole are not localised in the same region, we
have a group of low oscillator-strength states just above the ground
state.
\item That the group of low oscillator-strength states is
followed by a number of states with large oscillator strength which
are 2D excitonic states scattered on the T-shaped intersection.
\end{itemize}

We have also performed a detailed study of the exciton binding and
confinement energies as a function of the well width and Al molar
fraction for symmetric and asymmetric wires. We have shown
\begin{itemize}
\item That the highest binding energy in any structure so far
constructed is calculated to be 16.5 meV which is much smaller than
previously thought.
\item That for optimised asymmetric wires,
the confinement energy is enhanced but the binding energy is slightly
lower with respect to those in symmetric wires.
\item That for GaAs/Al$_{x}$Ga$_{1-x}$As wires we have obtained an
upper limit for the binding energy of around 25 meV in a 10 {\AA} wide
GaAs/AlAs structure which suggests that other materials need to be
explored in order to achieve room temperature applications.
In$_{y}$Ga$_{1-y}$As/Al$_{x}$Ga$_{1-x}$As might be a good candidate.
\end{itemize}

Our calculations are being used to design improved excitonic lasers
which will operate at room temperature. In particular we have studied
in detail the GaAs/Al$_{x}$Ga$_{1-x}$As structures in order to optimise
the confinement energy of the wire. 

We have described an experimental observation of an interesting phenomenon
of a two-mode lasing in asymmetric T-shaped wires. We have then
\begin{itemize} 
\item Developed a rate equation model for a two-mode
laser. Assuming that the laser action takes place for two different
exciton-like states, we can reproduce the same qualitative behaviour of
the laser, as observed in the experiment.
\item Applied the detailed numerical calculations for structures used in this
experiment and identified the origin of the two laser modes. The first
lasing line comes from the ground state exciton while the second line
is most probably associated with quasi-2D-exciton-like states strongly
perturbed by the quantum wire potential
\end{itemize}

\section{Future Directions}

\begin{itemize}
\item Our calculations are being used to design quantum wire lasers which
would operate at room temperature. This work, in collaboration with
Bell-Laboratories, is still in progress and detailed calculations
for structures with different materials or shapes might be useful.
\item It would be of interest to perform the rate equation calculations
for the exact parameters of the structures used in the experiment.
\item More complex extensions could consider doped quantum wires or
including a finite concentration of electron-hole pairs in the wire.
\end{itemize}

\appendix
\chapter{Convergence of the T-shaped Wire Calculations}
\chset{Convergence of the T-shaped Wire Calculations}
\label{app1}

\begin{table}[htbp]
\caption{Convergence of the energy (E) in meV and the oscillator
strength (OS) for states: 1st, 2nd, and 25th in the symmetric 70 {\AA}
quantum wire (as marked in Fig. \ref{symspec}, \ref{sym70st1} and
\ref{sym70st2}) with respect to the size of the basis set used in
diagonalisation.}
\begin{center}
\begin{tabular}{lcccccccc}
\multicolumn{9}{l}{} \\
\hline
\hline
\multicolumn{9}{l}{} \\
&\multicolumn{2}{c}{1st
}&
&\multicolumn{2}{c}{2nd
}&
&\multicolumn{2}{c}{25th
}
\\
Number of Basis Functions&E&OS&&E&OS&&E&OS\\
\hline
\multicolumn{9}{l}{} \\
10 x 10 x 10 &41.79&2.96&&48.72&0.98 \\
15 x 15 x 15 &41.44&3.66&&48.55&1.25&&63.34&2.42\\
17 x 17 x 17 &41.41&3.75&&48.54&1.28&&63.32&2.50\\ 
20 x 20 x 20 &41.36&3.92&&48.51&1.36&&63.29&2.62\\
21 x 21 x 21 &41.34&3.99&&48.51&1.38&&63.28&2.66 \\
\hline
\hline
\end{tabular}
\end{center}
\label{aptable1}
\end{table}

\begin{table}[htbp]
\caption{Convergence of the energy (E) in meV and the oscillator
strength (OS) for states: 1st, 2nd, and 25th in the symmetric 70 {\AA}
quantum wire (as marked in Fig. \ref{symspec}, \ref{sym70st1} and
\ref{sym70st2}) with respect to the number of points on the 2D grid.}
\begin{center} 
\begin{tabular}{lcccccccc}
\multicolumn{9}{l}{} \\
\hline
\hline
\multicolumn{9}{l}{} \\
&\multicolumn{2}{c}{1st
}&
&\multicolumn{2}{c}{2nd
}&
&\multicolumn{2}{c}{25th
}
\\
Number of Points&E&OS&&E&OS&&E&OS\\
\hline
\multicolumn{9}{l}{} \\
119 x 119
&41.44&3.66&&48.55&1.25&&63.34&2.42 \\
203 x 203 &41.62&3.57&&48.71&1.23&&63.54&2.31\\
245 x 245 &41.61&3.66&&48.73&1.24&&63.55&2.35\\ 
\hline
\hline
\end{tabular}
\end{center}
\label{aptable2}
\end{table}

\begin{figure}[htbp]
      \begin{center}
        \leavevmode
      \epsfxsize=14.0cm
      \epsfbox{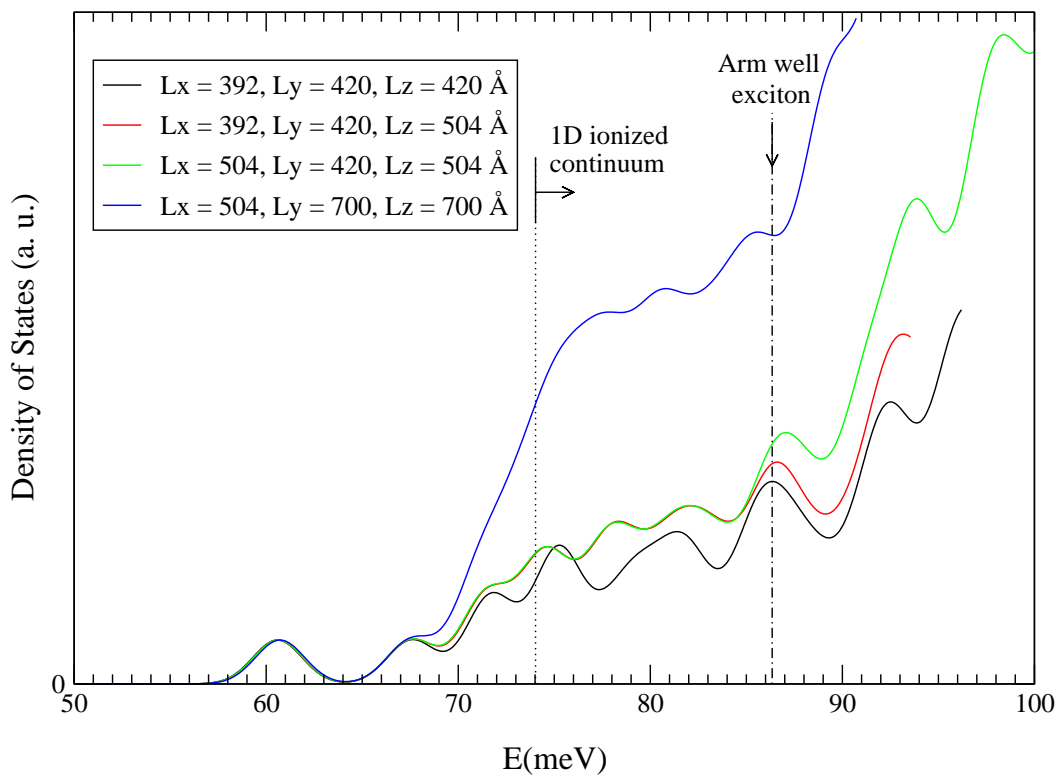}    
        \end{center}
\caption{Density of states in an asymmetric T-shaped
        structure with $D_x$ = 56 {\AA}, $D_y$ = 140 {\AA} discussed
        in Chapter \ref{bell} (Fig. \ref{bell3}) for four different unit
        cells. The line is produced by a Gaussian broadening with 1.5
        meV HWHM of calculated discrete energies. Because our system
        is finite we obtain only a sampling of the continuum
        states. When we increase the unit cell size we automatically
        calculate more states within the same energy region thus the
        density of states, for energies above the onset of continuum
        states, grows. However, the features of the curves remain the
        same, suggesting that the energies are well converged with
        respect to the unit cell size.}
\label{fig:ap_dens}
\end{figure}

\begin{figure}[htbp]
      \begin{center}
        \leavevmode
      \epsfxsize=13.5cm
      \epsfbox{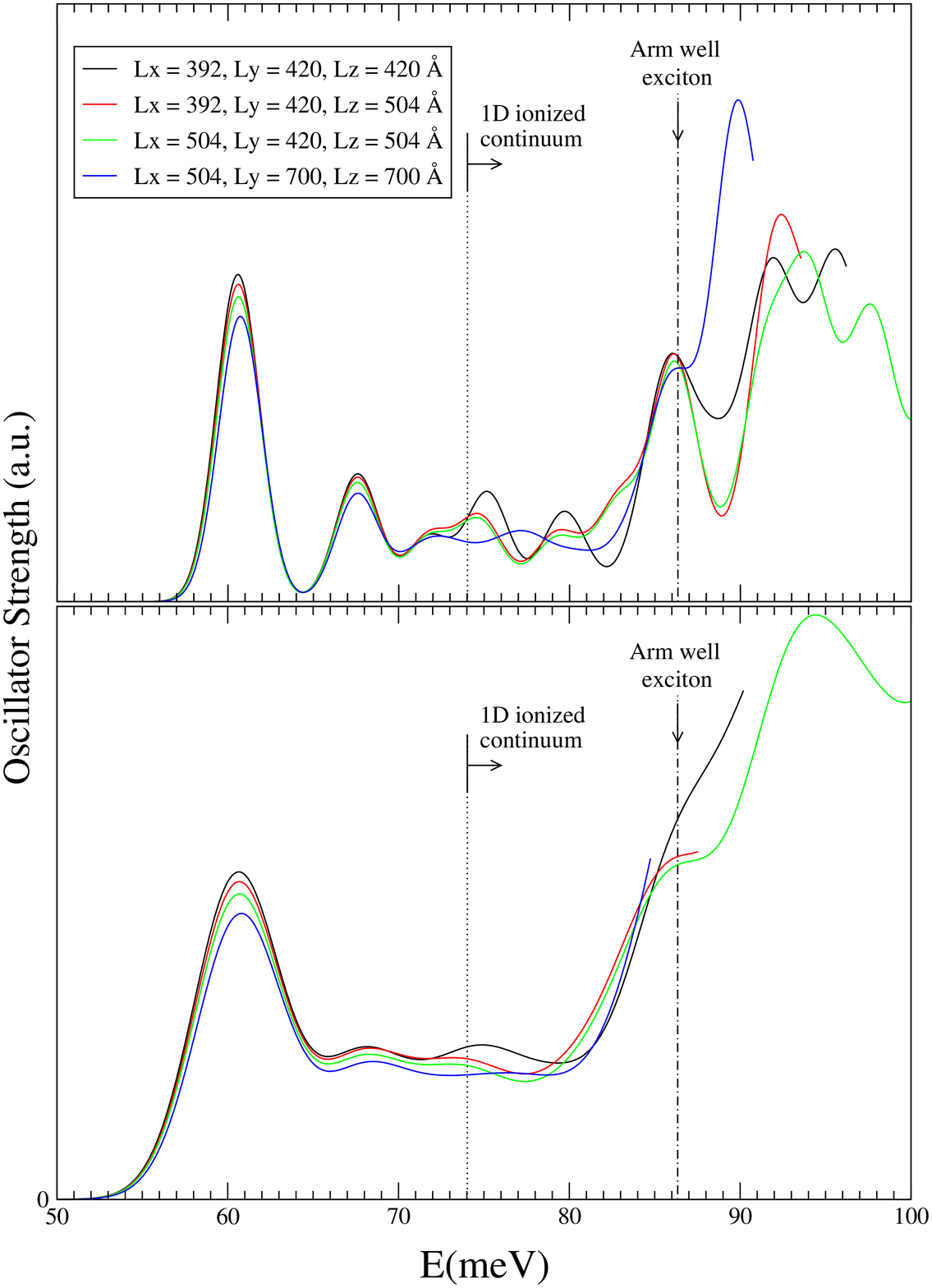}   
        \end{center}
\caption{Oscillator strength versus energy in an asymmetric T-shaped
        structure with $D_x$ = 56 {\AA}, $D_y$ = 140 {\AA}, discussed
        in Chapter \ref{bell} (Fig. \ref{bell3}), for four different unit
        cells. The line is produced by a Gaussian broadening with 1.5
        meV (upper panel) and 3 meV (lower panel) HWHM of calculated
        discrete states.}
\label{fig:ap_os}
\end{figure}

\chapter{Comments on the Effective Mass Approximation}
\chset{Comments on the Effective Mass Approximation}
\label{app2}
 
Calculations of the T-shaped wire states are performed using the
effective mass approximation with an anisotropic hole mass. In the
effective mass approximation electron and heavy hall single parabolic
bands are assumed. This approximation is justified in the energy
region of interest. 

\begin{figure}[htbp]
	\begin{center} \leavevmode \epsfxsize=14.5cm
	\epsfbox{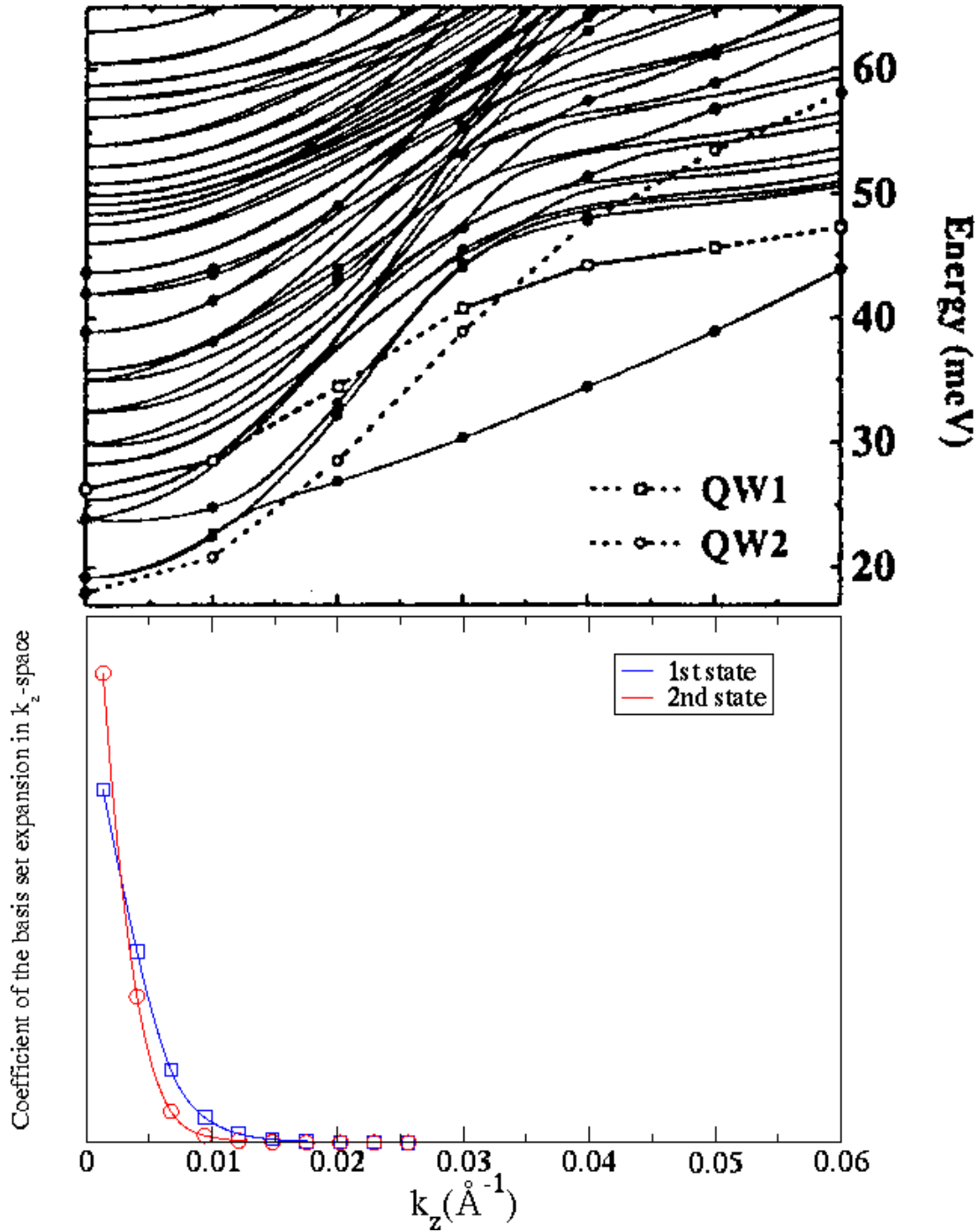}
	\end{center} \caption{Upper panel: Hole band structure along
	the z axis of the T-QWR. The full circles identify the states
	that are quasi-one-dimensional wire-like states or
	resonances. Dashed lines show the lowest-hole subbands of the
	isolated Arm and Stem quantum wells. Taken from
	\cite{molli}. Lower panel: Coefficient of the basis set
	expansion in $k$-space as a function of k in the z direction
	in the present calculations using the effective mass
	approximation.}
\label{fig:band} 
\end{figure}

In Figure \ref{fig:band} (upper panel) we present the hole band
structure along the wire direction, taken from \cite{molli}, for sample
S2 (see Table \ref{table}) experimentally studied in \cite{S1S2},
which consist of 53 {\AA} Stem and 48 {\AA} Arm quantum wells with
AlAs barriers. For the same sample, in Figure \ref{fig:band} (lower
panel), we present the contribution from different $k_z$ states in a
ground and a first excited state wavefunction given by equation
(\ref{basis-set}).  $\sum_{i,j}|c_{i,j,k_z}|^2$ is plotted as a
function of $k_z$.  Notice, that the hole bands are parabolic for
$k_z$ up to around 0.013 {\AA}$^{-1}$. The contribution from $k_z$
higher than 0.013 {\AA}$^{-1}$ is negligible for presented first two
states. The higher states are more spatially extended than the first
two states and thus they will decay in the $k_z$ space even faster
than the presented two states. The conduction band is shown to be
parabolic for $k_z$ up to 0.04 {\AA}$^{-1}$ for AlAs barriers
\cite{cond_band}. For lower Al concentration the conduction band is 
parabolic even for higher $k_z$ up to 0.11 {\AA}$^{-1}$ for Al
concentration, x=0.3, while the spatial extension of states would be
larger, and thus the region in $k$-space smaller than in the AlAs case.
Thus the parabolic band effective mass approximation is justified for
all structures being considered in this work.

Another issue is that the second hole band in Figure \ref{fig:band}
(upper panel), although parabolic in the energy region of interest,
splits into two branches with different, by around factor of two,
effective masses. The binding energies are, however, not that
sensitive to the hole effective mass. The binding energy of the light
hole exciton with the hole mass around eight times smaller than the
heavy hole mass differs only by 4 meV from the binding energy of the
heavy hole exciton. Pfeiffer {\it et al} have shown that the electron
and hole confinement energies calculated using the full eight band
$\vec{k} \cdot \vec{p}$ method agree within about 1 meV with those
obtained using the one band effective mass approximation
\cite{calc}. In the presence of the Coulomb interactions the effect on
the binding energy will be even smaller than that. Notice, that the
binding energy in these structures is a function of the confinement
energy (Fig. \ref{symEcEb}) with the changes in the confinement energy
corresponding to on average six times smaller changes in the binding
energy. Thus the difference in energies between the full eight band
$\vec{k} \cdot \vec{p}$ methods and our calculations would be smaller
than 1 meV for structures studied in this work.

\addcontentsline{toc}{chapter}{Bibliography}
\bibliographystyle{acm}

\end{document}